\documentclass[aps,prd,superscriptaddress,twoside,twocolumn,nofootinbib,10pt,%
showpacs,floatfix]{revtex4-1}

\usepackage{amsmath,amssymb}
\usepackage{dcolumn}
\usepackage{ulem} 
\usepackage{float}
\usepackage[usenames]{color}

\allowdisplaybreaks

\begin{document}

\title{Chiral effective theories from holographic QCD with scalars} 

\author{Masayasu Harada}
\email{harada@hken.phys.nagoya-u.ac.jp}
\affiliation{Department of Physics,  Nagoya University, Nagoya, 464-8602, Japan}

\author{Yong-Liang Ma}
\email{yongliangma@jlu.edu.cn}
\affiliation{College of Physics, Jilin University, Changchun, 130012, China}
\affiliation{Department of Physics,  Nagoya University, Nagoya, 464-8602, Japan}
    
\author{Shinya Matsuzaki}
\email{synya@hken.phys.nagoya-u.ac.jp}
\affiliation{Institute for Advanced Research, Nagoya University, Nagoya 464-8602, Japan.}
\affiliation{Department of Physics,  Nagoya University, Nagoya, 464-8602, Japan}

\date{\today}

\begin{abstract}
We develop a method for integrating out the heavy Kaluza-Klein modes of scalar type
 as well as those of vector and axial-vector types, in a class of hard-wall bottom-up
 approaches of holographic QCD models, including the Dirac-Born-Infeld and Chern-Simons parts. 
By keeping only the lowest-lying vector mesons, we first obtain an effective chiral 
Lagrangian of the vector mesons based on the hidden local symmetry, and 
all the low-energy constants in the HLS Lagrangian are expressed in terms of holographic integrals and, 
consequently, are fully determined by the holographic geometry and a few constants of mesons. 
We find that the Gell-Mann--Oakes--Renner relation is manifestly reproduced at the lowest order of derivative expansion. 
We also explicitly show that a naive inclusion of the Chern-Simons term 
 cannot reproduce the desired chiral anomaly in QCD, and hence, some counterterms should be provided: 
This implies that the holographic QCD models of hard-wall type cannot give definite predictions for the intrinsic parity-odd vertices involving vector and axial-vector mesons. After integrating out the vector mesons from the HLS Lagrangian,
we further obtain the Lagrangian of chiral perturbation theory for pseudoscalar mesons with all the low-energy constants fully determined.

\end{abstract}

\pacs{
11.25.Tq, 11.30.Rd, 12.39.Fe}
\maketitle

\section{Introduction}

Holographic QCD (hQCD), based on the gauge/gravity duality~\cite{Maldacena:1997re,Witten:1998qj}, has provided a novel approach to 
the low-energy dynamics of the strong interaction~\cite{Sakai:2004cn,Sakai:2005yt,Erlich:2005qh,Da Rold:2005zs}(for a recent review, see, e.g.,~\cite{Kim:2012ey}). 
In such an approach, the hQCD action is often written in five-dimensional anti-de Sitter space (AdS), so it is called AdS/QCD.  
The five-dimensional bulk fields, which satisfy the classical equations of motion, at the ultraviolet (UV) boundary serve as the sources of the QCD generating functionals of the connected Green functions in four-dimensional space-time, 
while their infrared (IR) boundary conditions (BCs) give rise to the discrete and infinite Kaluza-Klein (KK) tower of normalizable modes,
which correspond to the infinite number of hadrons of QCD in the large $N_c$ limit.

In contrast to the large $N_c$ physics expanded in terms of an infinite tower of hadrons, 
the low-energy dynamics of real-life QCD should be described by some specific effective theories,  
in which the dynamical degrees of freedom, appropriate only to the low-energy region, are included, 
with the effects from heavier resonances fully integrated out. For example, for describing the mesonic physics below $\mathcal{O}(100)$~MeV, 
it is sufficient to take into account only pseudoscalar mesons, arising as the pseudo Nambu-Goldstone bosons (NGBs) associated with 
the spontaneous breaking of chiral symmetry. However, if the interesting scale is at $\mathcal{O}(1)$~GeV, 
the lowest-lying vector mesons, in addition to the pseudoscalar mesons, should be present in the effective theory.

The purpose of this paper is to make a direct link between the hQCD, including an infinite number of QCD hadrons 
and low-energy effective theories described only by low-lying mesons in four-dimensional space-time, 
such as chiral perturbation theory (ChPT) of pseudoscalar  mesons~\cite{Wei:79} and chiral effective theory of vector mesons based on the hidden local symmetry (HLS), which includes the vector mesons in addition to the pseudoscalar mesons~\cite{Bando:1984ej}  
(for reviews, see Refs.~\cite{Bando:1987br,Harada:2003jx}).

Actually, in Refs.~\cite{Harada:2006di,Harada:2010cn}, the authors proposed a gauge-invariant methodology for integrating out the infinite tower of vector and axial-vector fields arising from hQCD models of the top-down approach   
such as the Sakai-Sugimoto model~\cite{Sakai:2004cn,Sakai:2005yt}, without bulk scalar fields, by regarding an infinite tower of the vector and axial-vector mesons in the hQCD model as a set of gauge bosons of HLSs~\cite{Son:2003et}.
The hQCD models were then reformulated in terms of the ChPT having the HLS (HLS-ChPT)~\cite{Tanabashi,Harada:2003jx} 
described by the NGBs, and the $\rho$ meson and its flavor partners.

In the present study, as an extension of the work in Refs.~\cite{Harada:2006di,Harada:2010cn}, 
we propose a gauge-invariant procedure for integrating out the infinite towers of
scalar and pseudoscalar fields, in addition to vector and axial-vector fields,  
from hQCD models: The resultant effective theory is reformulated as the HLS-ChPT. 
To be concrete, we employ bottom-up hard-wall-type of models 
such as those discussed in Refs.~\cite{Erlich:2005qh,Da Rold:2005zs} for QCD 
and \cite{Haba:2010hu} for applications to both QCD and other strongly coupled gauge dynamics.  
Our method is also applicable also to other types of hQCD models.

Compared to the Sakai-Sugimoto model, the most remarkable feature in  
models with bulk scalars is the inclusion of the effect of the explicit breaking of chiral symmetry via the current quark masses, which is  
encoded in the vacuum expectation value (VEV) of the bulk scalar. 
Thus, the method presented in this paper is highly nontrivial and valuable and  
makes it more straightforward to give holographic predictions in terms of the well-established chiral 
effective theories, ChPT or HLS-ChPT, including the effect from the explicit breaking of chiral symmetry.

After integrating out the heavy KK modes from hQCD models and keeping only the NGBs and the lowest-lying vector mesons,  
we arrive at the chiral effective theory of pions and vector mesons, HLS-ChPT, up to the ${\cal O}(p^4)$ terms,  
based on the derivative expansion which follows from the appropriate chiral-order counting rule. One thus automatically reproduces the Gell-Mann$-$Oakes$-$Renner (GOR) relation from the ${\cal O}(p^2)$ terms of the HLS-ChPT reduced from 
hQCD models, which are appropriately affected by the ${\cal O}(p^4)$ terms as in the case of 
the ChPT/HLS-ChPT.  The low-energy constants of the HLS-ChPT are actually expressed in terms of a few intrinsic quantities in hQCD models,  
which can be controlled by the equations of motion of the five-dimensional fields and some empirical values. 
Therefore, our procedure will provide a systematic way to estimate the low-energy constants of HLS-ChPT and consequently 
compute some physical quantities by using the HLS Lagrangian, including all the contributing terms up to $\mathcal{O}(p^4)$.

We also find another remarkable observation regarding the intrinsic parity (IP)-odd sector at the ${\cal O}(p^4)$ level: 
Because of the presence of the IR boundary in the case of hard-wall-type models, 
not only the Chern-Simons (CS) term but also certain counterterms  have to be taken into account 
in order to correctly reproduce the desired chiral anomaly in QCD. As one concrete sample, we present an explicit form of the counterterms, which, however, does not turn out to be a unique choice.

By integrating out the KK modes, keeping only the $\rho$ meson (and its flavor partners),  
the CS term (plus the counterterm) is reduced to the IP-odd part of the HLS-ChPT,  
which is constructed from the (covariantized) Wess-Zumino-Witten term~\cite{Wess:1971yu} plus  
HLS/chiral-invariant terms~\cite{Fujiwara:1984mp}~\footnote{
Similar invariant terms were discussed in \cite{Kaymakcalan:1984bz} based on a different treatment of $\rho$ meson field. }. 
The coefficients of the HLS/chiral-invariant terms are then represented fully in terms of the holographic integrals, 
though those coefficients are not uniquely fixed by the hQCD model itself because of the ambiguity in the choice of the counterterm: 
The hQCD models of hard-wall type cannot give definite predictions for 
IP-odd vertices involving vector and axial-vector mesons.

We further integrate out the lowest-lying vector mesons from the HLS Lagrangian to reach 
the ChPT, including terms up to $\mathcal{O}(p^4)$~\cite{Gasser:1983yg,Gasser:1984gg}. 
By this procedure, the low-energy constants of the ChPT Lagrangian are expressed 
 in terms of the parameters of the  hQCD models. 
We then find that the low-energy constants $L_9$ and $L_{10}$ in the ChPT Lagrangian satisfy $L_9 + L_{10}=0$, 
leading to a vanishing pion axial-vector form factor, denoted as $F_A$ by the Particle Data Group~\cite{Beringer:1900zz}.

The exploration of chiral effective theories from bottom-up hQCD models has so far been performed by several authors, 
such as those in Refs.~\cite{Hirn:2005nr,Colangelo:2012ipa}. 
In these analyses, the bulk scalar fields are not present so the effect of the explicit breaking of chiral symmetry via the current quark masses is not included. As for the IP-odd terms, in Ref.~\cite{Colangelo:2012ipa} the CS term has been discussed in the bottom-up hard-wall model;
however, the effect of the IR BC was not pointed out. 
In Ref.~\cite{Grigoryan:2008up}, the authors argued that, to reproduce the exact IP-odd action responsible for 
the $\pi\gamma^\ast\gamma^\ast$ coupling in four dimensions, 
a counterterm should be added to the original CS term to eliminate the surface term at the IR boundary. 
In contrast, in the present work we present a complete expression for the counterterms, including the IP-odd part written in terms of five-dimensional gauge fields, not restricted to specific terms like $\pi\gamma^\ast\gamma^\ast$, though the IP-odd part involving vector and axial-vector mesons is not uniquely fixed.

This paper is organized as follows: In Sec.~\ref{sec:hlsemergency} we explicitly show how the chiral symmetry and HLS are 
incorporated in hQCD models and then perform the KK decomposition for the bulk fields so as to 
reduce the hQCD model to four-dimensional gauge theory, including the infinite tower of the KK modes.   
In Sec.~\ref{sec:dbihls} we derive the IP-even part of the HLS-ChPT from the Dirac-Born-Infeld (DBI) part of the hQCD model. 
In Sec.~\ref{sec:cshls} the IP-odd part is also integrated out from the CS term, and the necessity of the counterterms for the IR value of the CS term is proposed. In Sec.~\ref{sec:chpt} the reduction to the ChPT is performed, and the implications of the low-energy constants of the ChPT Lagrangian are discussed. Our summary is given in the last section.

\section{Preliminary for the integrating-out procedure}  

\label{sec:hlsemergency}

Let us start from certain types of hQCD models, such as the ones proposed in Refs.~\cite{Erlich:2005qh,Da Rold:2005zs} 
and their modifications~\cite{Shock:2006qy,Shock:2006gt,Hirn:2005vk,Haba:2010hu}. 
These models are constructed in five dimensions with geometry $ds^2 = a^2(z) (\eta_{\mu\nu} dx^\mu dx^\nu - dz^2)$ 
where the fifth direction $z$ is confined in a finite interval $z_{\rm UV} <z< z_{\rm IR}$ (hard-wall type) 
and $a(z)$ is the warping factor in five dimensions.

The field content in the hQCD model  includes the gauge fields $L_M$ and $R_M$ $(M=\mu, z)$ 
for the  local $U(N_f)_L \times U(N_f)_R$ flavor symmetry in five dimensions, with $N_f$ being the number of flavors,  
and a bulk scalar field $\Phi_S$, which is dual to the chiral condensate. Generally, the model has the form constructed from the DBI and CS parts: 
\begin{eqnarray}
S & = & S_{\rm DBI} + S_{\rm CS} 
\,. 
\label{eq:hqcdaction}
\end{eqnarray}

Up to the quadratic order of the bulk fields, a generic form of the gauge-invariant DBI part
is written as
\begin{eqnarray}
S_{\rm DBI} & = & \frac{1}{g_5^2} 
\int d^4 x 
\int_{z_{\rm UV}}^{z_{\rm IR}} dz 
\sqrt{g} 
\nonumber \\ 
&& 
\times 
\Bigg(
 - \frac{1}{4} {\rm tr}\left[L_{MN} L^{MN} + R_{MN} R^{MN}\right] 
 \nonumber \\ 
&& + {\rm tr}\left[ D_M \Phi_S^\dag D^M \Phi_S\right] 
- m_{\Phi_S}^2 {\rm tr}\left[ \Phi_S^\dag \Phi_S\right] 
 \Bigg)
 \,, \label{eq:DBIpart}
\end{eqnarray}
where $1/g_5^2$ denotes the gauge coupling in five dimensions and 
$g={\rm det} \,g_{MN}=a^{10}(z)$. 
In Eq.~\eqref{eq:DBIpart}, the covariant derivative is defined as 
\begin{eqnarray} 
 D_M \Phi_S & = & 
 \partial_M \Phi_S - i L_M \Phi_S + i \Phi_S R_M 
 \,,
\end{eqnarray}
 and the field strength tensor is
\begin{eqnarray} 
  L_{MN} & = & \partial_M L_N - \partial_N L_M - i \left[L_M, L_N\right] 
\, ,
\end{eqnarray}
and similarly for $R_{MN}$.

According to the holographic dictionary, the bulk scalar mass $m_{\Phi_S}$ is related to the conformal dimension $\Delta$ of the corresponding operator 
in the dual gauge theory through~\cite{Gubser:1998bc,Witten:1998qj}
\begin{eqnarray}
m_{\Phi_S}^2 & = & - \frac{\Delta (4-\Delta)}{L^2} 
\,, 
\end{eqnarray}
with $L$ being some characteristic length scale of the holographic model, e.g., the AdS curvature radius in the case of the AdS background.

The CS term is written,  in terms of differential forms, as 
\begin{eqnarray}
S_{\rm CS} 
= \frac{N_c}{24\pi^2} \int_{M^4} \int_{z_{\rm UV}}^{z_{\rm IR}}  
\left[ \omega_5(R) - \omega_5(L) \right] 
\,,  \label{eq:CSpart}
\end{eqnarray}
 where $M^4$ denotes the manifold of the four-dimensional Minkowski space and 
\begin{eqnarray} 
 \omega_5(V) &=& 
{\rm tr}\left[V F_V^2 + \frac{i}{2} V^3 F_V - \frac{1}{10} V^5\right] 
 \,, \nonumber \\ 
F_V &=& dV - iV^2 
\,,  
\end{eqnarray} 
for $V=L,R$.

\subsection{Emergence of chiral symmetry}

First we explicitly show how the chiral symmetry of QCD is 
 realized in the hQCD models.

We begin by choosing the  following UV 
boundary condition
(BC) for the five-dimensional left- and right-gauge fields\footnote{
Note that this BC does not contradict the holographic recipe since, in the QCD generating functional, the external sources are introduced as the gauge fields of local chiral symmetry. 
}:
\begin{eqnarray}
L_M \Bigg|_{z=z_{\rm UV}} = R_M \Bigg|_{z=z_{\rm UV}} = 0 
\,. \label{eq:LRUVBC}
\end{eqnarray}
This boundary condition is not changed by the gauge transformation with respect to $g_{L,R}(x,z)$ satisfying  
\begin{eqnarray} 
  \partial_M g_L(x,z)\Bigg|_{z=z_{\rm UV}} = \partial_M g_R(x,z)\Bigg|_{z=z_{\rm UV}} = 0 
  \,, \label{eq:BCglobal}
\end{eqnarray} 
which can be checked by noting the gauge transformation laws for the five-dimensional gauge fields
\begin{eqnarray} 
 L_M & \to &  g_{L}(x,z) L_M g_{L}^\dag (x,z)  + i g_{L}(x,z) \partial_M g_{L}^\dag(x,z) \,, \nonumber\\
 R_M & \to &  g_{R}(x,z) R_M g_{R}^\dag (x,z) + i g_{R}(x,z) \partial_M g_{R}^\dag(x,z) 
\,. \nonumber \\ 
\label{L:R:5dtrans}
\end{eqnarray} 
This implies that there is a global symmetry at the UV boundary, which is identified with chiral symmetry in four dimensions.
Thus, chiral symmetry emerges from generic hQCD models.

In the action~\eqref{eq:hqcdaction}, the bulk scalar field $\Phi_S$ transforms 
under the five-dimensional gauge symmetry as 
\begin{eqnarray}
 \Phi_S \to g_L(x,z) \Phi_S g_R^\dag(x,z) \,,
 \label{phiS:chiral}
\end{eqnarray}
so that the UV boundary field of $\Phi_S$ also transforms under the residual chiral symmetry in four dimensions associated with $g_{L,R}(x,z)$ at the UV boundary. 
The chiral symmetry is thus spontaneously broken down to the diagonal vector $U(N_f)_V$ symmetry when 
$\Phi_S$ acquires a nonzero VEV, and hence the associated NGBs emerge in the model.

\subsection{Manifest realization of hidden local symmetry }

Next we show that HLS, as well as chiral symmetry, actually emerges in the hQCD models. For that purpose, it is convenient to work in the gauge
\begin{eqnarray}
L_z^g & = & R_z^g = 0\, ,
\label{eq:gaugefixLR}
\end{eqnarray}
where the superscript $g$ specifies the fields after the gauge fixing. 
This gauge fixing can be realized by a gauge transformation with respect to the transformation matrices [see Eq.(\ref{L:R:5dtrans})],
\begin{eqnarray} 
g_L^g(x,z) & = &  
{\rm P} \exp \left[  i \int^{z_{\rm IR}}_z d z' L_z(x,z') \right] 
\,, \nonumber\\ 
g_R^g(x,z) &=& 
{\rm P} \exp \left[ i \int^{z_{\rm IR}}_z d z' R_z(x,z') \right] 
\,, 
\label{eq:gaugez0}
\end{eqnarray}
where the symbol P denotes the path-ordered product. It should be noted that, 
although the five-dimensional gauge is now completely fixed, the action is still invariant under four-dimensional 
$U(N_f)_L\times U(N_f)_R$ gauge transformations 
which do not affect the gauge fixing in Eq.~\eqref{eq:gaugefixLR}.  
By writing the four-dimensional transformation matrices as $\tilde{g}_{L,R}(x) = g_{L,R}(x,z=z_{\rm IR})$, 
the gauge fields $L_\mu^g$ and $R_\mu^g$ and scalar $\Phi_S^g$ transform under this 
residual $U(N_f)_L\times U(N_f)_R$ symmetry 
as 
\begin{eqnarray} 
 L_\mu^g & \to &  \tilde{g}_{L}(x) L_\mu^g \tilde{g}_{L}^\dag (x)  + i \tilde{g}_{L}(x) \partial_\mu \tilde{g}_{L}^\dag(x) \,, \nonumber\\
 R_\mu^g & \to &  \tilde{g}_{R}(x) R_\mu^g \tilde{g}_{R}^\dag (x) + i \tilde{g}_{R}(x) \partial_\mu \tilde{g}_{R}^\dag(x) \,, \nonumber \\
 \Phi_S^g & \to & \tilde{g}_L(x) \Phi_S^g \tilde{g}_R^\dag(x) 
\,.   \label{PhiS:HLS:trans0}
 \end{eqnarray}
This residual $U(N_f)_L\times U(N_f)_R$ symmetry is nothing but the (generalized) HLS. 
Thus, the HLS has manifestly emerged from the hQCD models, 
while chiral symmetry appears to be ``hidden" in Eq.~\eqref{PhiS:HLS:trans0}:  
Looking at the UV boundary values for the left- and right-gauge fields, 
\begin{eqnarray} 
 L_\mu^g \Bigg|_{z=z_{\rm UV}} &=& i \xi^g_L \partial_\mu \xi_L^{g\dag} 
\,, \quad
 R_\mu^g \Bigg|_{z=z_{\rm UV}} = i \xi_R^g \partial_\mu \xi_R^{g\dag} 
\,, \label{eq:UVBCg}
\end{eqnarray}
where 
\begin{eqnarray} 
\xi_{L}^g(x) & = &  
{\rm P} \exp \left[  i \int^{z_{\rm IR}}_{z_{\rm UV}} d z' L_z(x,z') \right] 
\,, \nonumber\\
\xi_{R}^g(x) & = &
{\rm P} \exp \left[ i \int^{z_{\rm IR}}_{z_{\rm UV}} d z' R_z(x,z') \right] 
\,, 
\label{eq:gaugez1}
\end{eqnarray}
one finds that these fields transform as
\begin{eqnarray} 
\xi_{L}^g(x) & \to & \tilde{g}_L(x)\xi_{L}^g(x) g_L^\dag, 
\nonumber \\ 
\xi_{R}^g(x) &\to& \tilde{g}_R(x)\xi_{R}^g(x) g_R^\dag
\,, 
\end{eqnarray}
in which $g_L$ and $g_R$ denote the transformation matrices regarding the global chiral $U(N_f)_L\times U(N_f)_R$ symmetry.

We further fix the gauge degree of freedom corresponding to the axial part of $\tilde{g}_L$ and $\tilde{g}_R$ to keep only the vector part, $h(x)$, of $\tilde{g}_L$ and $\tilde{g}_R$\footnote{
By using $L_\mu^{g}, R_\mu^{g}$ and $ \Phi_S^g$ in Eq.(\ref{PhiS:HLS:trans0}), after KK mode expansion as discussed below, 
one could straightforwardly derive the chiral effective theory based on the generalized HLS approach. 
Instead of proceeding along such a generalized HLS, in the present study we focus on reduction of hQCD models to the HLS.    
}. 
After this gauge fixing we express the bulk fields $L_\mu^{g}, R_\mu^{g}, \Phi_S^{g}$, $\xi_L^g$ and $\xi_R^g$ in Eq.(\ref{eq:gaugez1}) as $L_\mu^{gg}, R_\mu^{gg}, \Phi_S^{gg}, \xi_L^{gg}$ and $\xi_R^{gg}$, respectively.  
We then find that these fields transform under the residual vectorial HLS regarding $h(x)$ as 
\begin{eqnarray} 
 L_\mu^{gg} & \to &  h(x) L_\mu^{gg} h^\dag (x)  + i h(x) \partial_\mu h^\dag(x) \,, \nonumber\\
 R_\mu^{gg} & \to &  h(x) R_\mu^{gg}h^\dag (x) + i h(x) \partial_\mu h^\dag(x) \,, \nonumber \\
 \Phi_S^{gg} & \to & h(x) \Phi_S^{gg} h^\dag(x) ,\nonumber\\
 \xi_{L}^{gg} & \to & h(x)\xi_{L}^{gg} g_L^\dag, \nonumber\\
 \xi_{R}^{gg} & \to & h(x)\xi_{R}^{gg} g_R^\dag 
\,.   \label{PhiS:HLS:trans1}
 \end{eqnarray}
Under this gauge fixing, the UV boundary conditions in Eq.~\eqref{eq:UVBCg} are changed to 
\begin{eqnarray} 
 L_\mu^{gg} \Bigg|_{z=z_{\rm UV}} &=& i \xi_L^{gg} \partial_\mu \xi_L^{gg\dag} 
\,, \nonumber \\ 
R_\mu^{gg} \Bigg|_{z=z_{\rm UV}} &=& i \xi_R^{gg} \partial_\mu \xi_R^{gg\dag} 
\,.  \label{UV:LR}
\end{eqnarray} 
 By using $\xi_L^{gg}$ and $\xi_R^{gg}$ one can construct the chiral field $U$ as
\begin{eqnarray}
U = \xi_L^{gg\dag} \cdot \xi_R^{gg} 
= e^{2i \pi/F_\pi}
\, ,
\label{eq:UxiLR}
\end{eqnarray}
where $\pi$ denotes the NGB fields and $F_\pi$ the associated decay constant. 
The chiral field $U$ transforms in the standard manner as 
\begin{eqnarray}
 U \to g_L U g_R^\dag 
 \,. \label{U:chiral}
\end{eqnarray} 
Thus, we can construct the chiral effective theory which is invariant under the HLS, $h(x)$, reduced from hQCD models.

\subsection{KK decomposition}

Let us next perform the KK decomposition for the bulk fields $L_\mu^{gg}, R_\mu^{gg}$ and $\Phi_S^{gg}$ in Eq.(\ref{PhiS:HLS:trans1}) 
to explicitly express the hQCD models in terms of fields in four dimensions which are invariant under both the chiral symmetry and the HLS.

Since the residual HLS is a vector gauge symmetry [$h(x)$], 
we may expand the left and right gauge fields $L_\mu^{gg}$ and $R_\mu^{gg}$ in terms of 
KK modes of vector and axial-vector gauge fields $V_\mu^{gg}= (R_\mu^{gg} + L_\mu^{gg})/2$ and $A_\mu^{gg}=(R_\mu^{gg} - L_\mu^{gg})/2$.  
From Eq.~\eqref{PhiS:HLS:trans1} we then see that $V_\mu^{gg}$ and $A_\mu^{gg}$ transform under the HLS as 
\begin{eqnarray} 
V_\mu^{gg}(x,z) & \to & h(x) V_\mu^{gg}(x,z) h^\dag(x) + i h(x) \partial_\mu h^\dag(x) 
\,, \nonumber \\ 
A_\mu^{gg}(x,z) & \to & h(x) A_\mu^{gg}(x,z) h^\dag(x)
\,. \label{Vmu:HLS:trans}
\end{eqnarray} 
Note that the vector field $V_\mu^{gg}$ transforms inhomogeneously, like gauge fields, 
under the HLS transformation, while the axial-vector field $A_\mu^{gg}$ transforms homogeneously, like ``matter" fields. 
The UV boundary values for $V_\mu^{gg}$ and $A_\mu^{gg}$ are then, by using Eq.~\eqref{UV:LR},  expressed as 
\begin{eqnarray} 
  V_\mu^{gg}  
\Bigg|_{z=z_{\rm UV}} =  \alpha_{\mu ||}^{gg} 
\,, \qquad 
A_\mu^{gg} \Bigg|_{z=z_{\rm UV}} = \alpha_{\mu \perp}^{gg}
\,,  \label{UV:VA}
\end{eqnarray}
where 
\begin{equation} 
\alpha_{\mu ||, \perp}^{gg} 
= \frac{1}{2i} \left( \partial_\mu \xi_R^{gg} \cdot \xi_R^{gg \dag} \pm  \partial_\mu \xi_L^{gg} \cdot \xi_L^{gg\dag} \right) 
\,, 
\end{equation} 
which transform under the HLS as
\begin{eqnarray} 
 \alpha_{\mu \parallel}^{gg} &\to& 
 h(x) \alpha_{\mu \parallel}^{gg} h^\dag(x) + i h(x) \partial_\mu h^\dag(x) 
 \,, \nonumber \\ 
  \alpha_{\mu \perp}^{gg} &\to& 
 h(x) \alpha_{\mu \perp}^{gg} h^\dag(x)
 \,. \label{alphapara:HLS:trans}
\end{eqnarray} 
The expressions of KK decomposition for $V_\mu^{gg}$ and $A_\mu^{gg}$ thus take the form
\begin{eqnarray}
V_\mu^{gg}(x,z) &=& \alpha_{\mu \parallel}^{gg}(x) \psi^V_0(z) + \sum_{n \ge 1} V_\mu^{gg(n)}(x) \psi_n^V(z) 
\,, \nonumber \\ 
A_\mu^{gg}(x,z) &=& \alpha_{\mu \perp}^{gg}(x) \psi_0^A(z) + \sum_{n \ge 1} A_\mu^{gg (n)}(x) \psi_n^A(z) 
\,, \label{KK:decompose}
\end{eqnarray} 
where a set of wave functions $\{\psi^V_0,\psi_0^A,\psi_n^V,\psi_n^A  \}$ has been introduced with the UV boundary conditions:  
\begin{eqnarray} 
\psi_0^V(z=z_{\rm UV}) & = & 
\psi_0^A(z=z_{\rm UV}) = 1 
\,, \nonumber \\ 
\psi_n^V(z=z_{\rm UV}) & = & 
\psi_n^A(z=z_{\rm UV}) = 0 
\,.  
\end{eqnarray}
 In Eq.~(\ref{KK:decompose}) we have introduced the massive KK mode fields $V_\mu^{gg (n)}$ and $A_\mu^{gg (n)}$ which transform under the HLS as 
\begin{eqnarray} 
V_\mu^{gg (n)}(x) &\to&  
h(x) V_\mu^{gg (n)}(x) h^\dag(x) + i h(x) \partial_\mu h^\dag(x) 
\,, \nonumber \\ 
A_\mu^{gg (n)}(x) & \to & 
h(x) A_\mu^{gg (n)}(x) h^\dag(x) 
 \,. \label{An:HLS:trans}
\end{eqnarray}

The zero mode wave function $\psi_0^A$ is determined by solving the eigenvalue equation derived from the action Eq.~\eqref{eq:DBIpart},
\begin{eqnarray} 
\partial_z (a(z) \partial_z \psi_0^A(z)) - 4 a^3(z) (\Phi_S^0(z))^2 \psi_0^A(z) = 0 
\,, \label{EOM:psi0A}
\end{eqnarray}
where $\Phi_S^0(z)$ is the VEV of the bulk scalar field $\Phi_S$ which will be given later. The eigenvalue equation \eqref{EOM:psi0A} could be solved by considering the BCs   
\begin{eqnarray} 
&& \psi_0^A(z=z_{\rm UV}) = 1 
  \,, \nonumber \\ 
&& 
\partial_z \psi_0^A(z) \Bigg|_{z=z_{\rm IR}}  =0 
\,,
\end{eqnarray}
in which the IR BC has been set so as to eliminate the IR value of the axial-vector fields in the action as done in Ref.~\cite{Da Rold:2005zs}. 
Similarly, from the action Eq.~\eqref{eq:DBIpart}, we obtain the eigenvalue equations for  
the massive KK mode wave functions $\{\psi_n^V, \psi_n^A  \}$
\begin{eqnarray} 
0 & = &  a^{-1}(z) \partial_z (a(z) \partial_z \psi_n^V(z)) + m_{V_n}^2 \psi_n^V(z)  
\,, \nonumber \\ 
0 & = &  a^{-1}(z) \partial_z (a(z) \partial_z \psi_n^A(z)) - 4 a^2(z) (\Phi_S^0(z))^2 \psi_n^A(z) \nonumber\\
& & {} + m_{A_n}^2 \psi_n^A(z)
\,, 
\end{eqnarray} 
with the BCs 
\begin{eqnarray} 
\psi_n^V(z=z_{\rm UV}) & = & 
\psi_n^A(z=z_{\rm UV}) = 0 
\,, \nonumber \\ 
\partial_z \psi_n^V(z) \Bigg|_{z=z_{\rm IR}}
& = & 
\partial_z \psi_n^A(z) \Bigg|_{z=z_{\rm IR}} =0 
\, ,
\end{eqnarray} 
where the vanishing IR boundary values for the massive KK fields have also been taken into account.

Note that the HLS transformation properties for $V_\mu^{gg}(x,z)$, $\alpha_{\mu \parallel}^{gg}$ and $V_\mu^{gg(n)}$ 
in Eqs.~\eqref{Vmu:HLS:trans}, \eqref{An:HLS:trans} and \eqref{alphapara:HLS:trans} constrain the 
wave functions $\psi_0^V$ and $\psi_n^V$ as follows: 
\begin{eqnarray}
 \psi_0^V(z) + \sum_{n\ge 1} \psi_n^V(z) = 1 
\,,
\end{eqnarray} 
which allows us to rewrite the KK decomposition form of $V_\mu(x,z)$ in Eq.~\eqref{KK:decompose} as 
\begin{eqnarray}
V_\mu^{gg}(x,z) = \alpha_{\mu ||}^{gg}(x)  + \sum_{n \ge 1} ( V_\mu^{gg(n)}(x) - \alpha_{\mu ||}^{gg}(x) ) \psi_n^V(z) . \nonumber\\
\label{Vmu:KK:decompose}
\end{eqnarray}

Let us turn to the KK decomposition for the scalar sector. The bulk scalar field $\Phi_S^{gg}$ can be parametrized by the scalar and pseudoscalar fields, $S^{gg}(x,z)$ and $P^{gg}(x,z)$, as 
\begin{eqnarray} 
\Phi_S^{gg}(x,z) = v_S(z) +  S^{gg}(x,z)  + iP^{gg}(x,z ) 
\, , \label{eq:paraSgg}
\end{eqnarray} 
where $v_S(z)$ denotes the VEV of the bulk scalar in the chiral limit, related to the chiral condensate, 
which corresponds to the normalizable solution of 
the equation of motion for $\Phi_S^0(z) = \langle \Phi_S(x,z) \rangle$: 
\begin{equation} 
  \partial_z \left( a^3 \partial_z \Phi_S^0 \right) - a^5 m_{\Phi_S}^2 \Phi_S^0 = 0 
  \,. \label{PhiS0} 
\end{equation} 
For simplicity, the $v_S(z)$ is assumed to be $N_f$-flavor diagonal, 
$\propto 1_{N_f \times N_f}$, and the explicit expression will be specified once the warp factor $a(z)$ is fixed. 
According to the holographic recipe, 
the non-normalizable solution of Eq.(\ref{PhiS0}), which we write as $u_S(z)$, 
serves as the source terms $(\chi_S, \chi_P)$ for the scalar and pseudoscalar currents at the UV boundary. 
The $u_S(z)$ is thus embedded in the UV BCs of 
the bulk fields $S^{gg}(x,z)$ and $P^{gg}(x,z)$ as follows:  
\begin{eqnarray} 
\frac{S^{gg}(x,z)}{u_S(z)} \Bigg|_{z=z_{\rm UV}} 
& = & \chi_S(x) \equiv \frac{\hat{\chi}(x) + \hat{\chi}^\dag (x)}{2} 
\,, \nonumber \\ 
\frac{P^{gg}(x,z)}{u_S(z)}  \Bigg|_{z=z_{\rm UV}} 
& = & \chi_P(x) \equiv \frac{\hat{\chi}(x) - \hat{\chi}^\dag (x)}{2}
\,.
\label{eq:BCSP}
\end{eqnarray}
When the warping factor $a(z)$ is chosen in such a way that the metric asymptotically goes like the AdS, 
one can see that $u_S(z)$ has the following UV asymptotic form:  
\begin{equation} 
  u_S(z) \sim
 \left( \frac{z}{L} \right)^{4-\Delta} 
\qquad 
{\rm as} 
\qquad 
z \to z_{\rm UV} (\to 0)
  \,. 
\end{equation}
Note that the UV boundary values $\chi_S$ and $\chi_P$ play the role of 
spurion fields corresponding to the sources of the scalar and pseudoscalar currents 
with scaling dimension $\Delta$, which can be expressed by 
${\cal M}$, the current-quark mass matrix accounting for the explicit breaking of the 
chiral symmetry, as  
\begin{eqnarray}
\chi_S 
   & = & 
\frac{\xi_L^{gg} {\cal M} \xi_R^{gg\dag} + \xi_R^{gg} {\cal M} \xi_L^{gg\dag} }{2}
\,, \nonumber \\ 
\chi_P 
&=& 
\frac{\xi_L^{gg} {\cal M} \xi_R^{gg\dag} - \xi_R^{gg} {\cal M} \xi_L^{gg\dag} }{2i}
\,. 
\end{eqnarray} 
In particular, 
the scalar spurion field $\chi_S$ can be expanded in terms of the current quark masses $m_q$ and the NGB fields as  
\begin{eqnarray} 
\chi_S 
&=& \frac{1}{2} \left( \xi_L^{gg} {\cal M} \xi_R^{gg\dag} + \xi_R^{gg} {\cal M} \xi_L^{gg\dag} \right)
\nonumber \\ 
&=& \left( 
 \begin{array}{cccc} 
 m_u & \\ 
  & m_d \\ 
  & & m_s \\ 
  & & & \ddots \\ 
\end{array}  
\right) 
+ {\cal O}(m_q \cdot \pi/F_\pi) 
\,. 
\end{eqnarray}

The $S^{gg}(x,z)$ and $P^{gg}(x,z)$ can thus be expanded in terms of their KK modes $S_n^{gg}(x)$ and $P_n^{gg}(x)$ as
\begin{eqnarray} 
S^{gg} (x,z) &=& 
u_S(z) \, \chi_S(x) +   
\sum_{n \ge 1} S_n^{gg}(x) \psi_n^S(z)  
\,, \nonumber \\ 
P^{gg} (x,z) &=& 
u_S(z) \, 
\chi_P(x) + 
\sum_{n \ge 1} P_n^{gg}(x) \psi^P_n(z) 
\,,     
\label{PhiS:expand} 
 \end{eqnarray} 
where $S_n^{gg}(x)=S^{gg A}_n(x) T^A  (P_n^{gg}(x)=P^{ggA}_n(x) T^A ) $, with $T^A$ 
($A=0,\cdots N_f^2-1$) being the $U(N_f)$ generators normalized as ${\rm tr}[T^AT^B]=\delta^{AB}/2$ with $T^0 = 1/\sqrt{2N_f}\cdot {\bf 1}_{N_f \times N_f}$. 
From Eq.~\eqref{eq:BCSP}, the wave functions $\psi_n^S$ and $\psi_n^P$ satisfy the normalizable UV BCs: 
\begin{equation} 
  \psi_n^S(z=z_{\rm UV}) = \psi_n^P(z=z_{\rm UV})=0 
  \,,
\end{equation}
and the IR BCs are to be determined when an IR boundary potential for the bulk scalar field is specified, 
for example, as in the model proposed in Ref.~\cite{DaRold:2005vr}.

\subsection{Gauging chiral symmetry}

The external gauge fields coupled to quark currents in QCD can be incorporated by gauging the global chiral symmetry. 
Because the UV boundary values of the five-dimensional bulk fields are set to the sources in QCD, 
introducing the external gauge fields just modifies the zero mode fields 
$\alpha_{\mu \parallel}^{gg}$ and $\alpha_{\mu \perp}^{gg}$ included in $V_\mu^{gg}(x,z)$ and $A_\mu^{gg}(x,z)$ 
in Eqs.~\eqref{Vmu:KK:decompose} and \eqref{KK:decompose} to their covariant forms
\begin{eqnarray} 
 \alpha_{\mu \parallel, \perp}^{gg} & \to & \alpha_{\mu \parallel, \perp}^{gg} 
 = 
\frac{{\cal D}_\mu \xi_R^{gg} \cdot \xi_R^{gg\dag} \pm {\cal D}_\mu \xi_L^{gg} \cdot \xi_L^{gg\dag} }{2i}
 \,, 
\end{eqnarray} 
where 
\begin{eqnarray} 
{\cal D}_\mu \xi_{R}^{gg} & = &\partial_\mu \xi_{R}^{gg} + i \xi_{R}^{gg} {\cal R}_\mu \,,
\nonumber \\  
{\cal D}_\mu \xi_{L}^{gg} 
&=& \partial_\mu \xi_{L}^{gg} + i \xi_{L}^{gg} {\cal L}_\mu 
\,, 
\end{eqnarray} 
with ${\cal R}_\mu$ and ${\cal L}_\mu$ being the external gauge fields for $U(N_f)_R$ and $U(N_f)_L$, respectively. 
The HLS transformation laws for these modified 1-forms do not change from those given in Eq.~\eqref{alphapara:HLS:trans}.

\section{The IP-even part of the hidden local symmetry Lagrangian}

\label{sec:dbihls}

We next explicitly derive the IP-even part of the HLS Lagrangian up to $\mathcal{O}(p^4)$ from the DBI part of the hQCD model in Eq.~\eqref{eq:DBIpart}. 
In the $L_z = R_z = 0$ gauge the DBI part in Eq.~\eqref{eq:DBIpart} is expressed 
in terms of the vector and axial-vector gauge fields $V_\mu^{gg}$ and $A_\mu^{gg}$ as
\begin{widetext}
\begin{eqnarray} 
 S_{\rm DBI} 
& = & 
\frac{1}{g_5^2} 
\int d^4 x 
\int_{z_{\rm UV}}^{z_{\rm IR}} dz 
\Big\{ 
-\frac{1}{2} a(z) 
{\rm tr}\left[V_{\mu\nu}^{gg} V^{gg\mu\nu} - 2 V_{\mu z}^{gg}V^{gg\mu z} + A_{\mu\nu}^{gg} A^{gg\mu\nu} - 2 A_{\mu z}^{gg} A^{gg\mu z}\right] \nonumber \\
& &\qquad\qquad\qquad\qquad\;{} + a^3(z) 
\left( 
{\rm tr}[ D_\mu \Phi_S^{gg\dag} D^\mu \Phi_S^{gg} -  \partial_z \Phi_S^{gg\dag} \partial_z \Phi_S^{gg} ] 
- a^2(z) m_{\Phi_S}^2 {\rm tr}[ \Phi_S^{gg\dag} \Phi_S^{gg}] 
 \right) 
\Big\} 
 \,, \label{DBI:action2}
\end{eqnarray}
\end{widetext}
where 
\begin{eqnarray} 
V_{\mu\nu}^{gg} & = & \partial_\mu V_\nu^{gg} - \partial_\nu V_\mu^{gg} - i [V_\mu^{gg}, V_\nu^{gg}] - i [A_\mu^{gg}, A_\nu^{gg}] 
\,, \nonumber \\ 
A_{\mu\nu}^{gg} & = & \partial_\mu A_\nu^{gg} - \partial_\nu A_\mu^{gg} - i [V_\mu^{gg}, A_\nu^{gg}] - i [A_\mu^{gg}, V_\nu^{gg}] 
\,, \nonumber \\ 
V_{\mu z}^{gg} & = & {} - \partial_z V_\mu^{gg} 
\,, \nonumber \\ 
A_{\mu z}^{gg} & = & {} - \partial_z A_\mu^{gg} 
\,, \nonumber \\ 
D_\mu \Phi_S^{gg} & = & \partial_\mu \Phi_S^{gg} - i [ V_\mu^{gg}, \Phi_S^{gg} ] + i \{ A_\mu^{gg}, \Phi_S^{gg} \} 
\,. 
\end{eqnarray} 
  
Following the gauge-invariant procedure proposed in Ref.~\cite{Harada:2010cn},  
we shall integrate out the KK fields to reduce the DBI action Eq.(\ref{DBI:action2}) to 
the HLS action in terms of the derivative expansion of the pseudoscalar meson and the lowest-lying vector meson fields up to chiral order $\mathcal{O}(p^4)$. The resultant form should be the same as that constructed from the general phenomenological consideration~\cite{Tanabashi,Harada:2003jx}  
with a single trace, which is consistent with the large $N_c$ limit\footnote{
The notation used in Ref.~\cite{Harada:2003jx} for terms including $\hat{\chi}$ can be recovered by taking $BF_\pi=BF_\chi$ 
and replacing $\hat{\chi}$ with $\hat{\chi}/(2B)$. }
\begin{eqnarray} 
{\cal L}_{\rm HLS} = {\cal L}_{\rm HLS}^{(2)} + {\cal L}_{\rm HLS}^{(2)\chi} + {\cal L}_{\rm HLS}^{(4)} + {\cal L}_{\rm HLS}^{(4)\chi} 
\,, \label{HLS:L2:L4} 
\end{eqnarray} 
with~\footnote{
 The definition of $\hat{\chi}$ given in Eq.(\ref{eq:BCSP})
has been changed from 
that in Ref.~\cite{Harada:2003jx} by factoring out the parameter $B$ and dividing by 2. 
}
\begin{widetext} 
\begin{eqnarray} 
{\cal L}_{\rm HLS}^{(2)} & = & F_{\pi}^2 {\rm tr}\left[\hat{\alpha}_{\mu\perp} \hat{\alpha}^{\mu}_{\perp}\right] + F_\sigma^2 {\rm tr}\left[\hat{\alpha}_{\mu \parallel} \hat{\alpha}^{\mu}_{\parallel} \right] - \frac{1}{2g^2}{\rm tr}\left[ V_{\mu\nu} V^{\mu\nu} \right] ,   \nonumber\\ 
 {\cal L}_{\rm HLS}^{(2)\chi} 
& = & B F_\pi^2 {\rm tr}[\chi_S], \nonumber\\ 
{\cal L}_{\rm HLS}^{(4)} & = &  
 y_1 \, {\rm tr}\left[{\hat{\alpha}}_{\mu\perp}{\hat{\alpha}}^{\mu}_{\perp}{\hat{\alpha}}_{\nu\perp}{\hat{\alpha}}^{\nu}_{\perp}\right]
+ y_2 \, {\rm tr}\left[{\hat{\alpha}}_{\mu\perp}{\hat{\alpha}}_{\nu\perp}{\hat{\alpha}}^{\mu}_{\perp}{\hat{\alpha}}^{\nu}_{\perp}\right]
+y_3 \, {\rm tr}\left[{\hat{\alpha}}_{\mu \parallel}{\hat{\alpha}}^{\mu}_{\parallel}
{\hat{\alpha}}_{\nu ||} {\hat{\alpha}}^{\nu}_{\parallel} \right] 
+y_4 \, {\rm tr}\left[{\hat{\alpha}}_{\mu \parallel}{\hat{\alpha}}_{\nu \parallel}
{\hat{\alpha}}^{\mu}_{\parallel}{\hat{\alpha}}^{\nu}_{\parallel}\right] \nonumber \\
& & {}
+y_5 \, {\rm tr}\left[{\hat{\alpha}}_{\mu\perp}{\hat{\alpha}}^{\mu}_{\perp}
{\hat{\alpha}}_{\nu \parallel}{\hat{\alpha}}^{\nu}_{\parallel}\right]
+y_6 \, {\rm tr}\left[{\hat{\alpha}}_{\mu\perp}{\hat{\alpha}}_{\nu\perp}
{\hat{\alpha}}^{\mu}_{\parallel}{\hat{\alpha}}^{\nu}_{\parallel}\right]
+y_7 \, {\rm tr}\left[{\hat{\alpha}}_{\mu\perp}{\hat{\alpha}}_{\nu\perp}
{\hat{\alpha}}^{\nu}_{\parallel}{\hat{\alpha}}^{\mu}_{\parallel}\right] \nonumber \\
& & {}
+y_8 \, \left\{ 
{\rm tr}\left[{\hat{\alpha}}_{\mu\perp}{\hat{\alpha}}^{\mu}_{\parallel}
{\hat{\alpha}}_{\nu\perp}{\hat{\alpha}}^{\nu}_{\parallel}\right]
+
{\rm tr}\left[{\hat{\alpha}}_{\mu\perp}{\hat{\alpha}}^{\nu}_{\parallel}
{\hat{\alpha}}_{\nu\perp}{\hat{\alpha}}^{\mu}_{\parallel}\right]
\right\}
+y_9 \, {\rm tr}\left[{\hat{\alpha}}_{\mu\perp}{\hat{\alpha}}^{\nu}_{\parallel}
{\hat{\alpha}}_{\mu\perp}{\hat{\alpha}}_{\parallel \nu} \right]
\nonumber \\
& & {}
+z_1 \, {\rm tr}\left[{\hat{\mathcal{V}}}_{\mu\nu}{\hat{\mathcal{V}}}^{\mu\nu}\right]
+z_2 \, {\rm tr}\left[{\hat{\mathcal{A}}}_{\mu\nu}{\hat{\mathcal{A}}}^{\mu\nu}\right]
+z_3 \, {\rm tr}\left[{\hat{\mathcal{V}}}_{\mu\nu}V^{\mu\nu}\right] 
+iz_4 \,{\rm tr}\left[V_{\mu\nu}{\hat{\alpha}}^{\mu}_{\perp}{\hat{\alpha}}^{\nu}_{\perp}\right]
+iz_5 \,{\rm tr}\left[V_{\mu\nu}{\hat{\alpha}}^{\mu}_{\parallel}
{\hat{\alpha}}^{\nu}_{\parallel}\right]
\nonumber \\
& & {}
+iz_6 \, {\rm tr}\left[{\hat{\mathcal{V}}}_{\mu\nu}
{\hat{\alpha}}^{\mu}_{\perp}{\hat{\alpha}}^{\nu}_{\perp}\right]
+iz_7 \, {\rm tr}\left[{\hat{\mathcal{V}}}_{\mu\nu}
{\hat{\alpha}}^{\mu}_{\parallel}{\hat{\alpha}}^{\nu}_{\parallel}\right]
-iz_8 \, {\rm tr}\left[{\hat{\mathcal{A}}}_{\mu\nu}
\left({\hat{\alpha}}^{\mu}_{\perp}{\hat{\alpha}}^{\nu}_{\parallel}
+{\hat{\alpha}}^{\mu}_{\parallel}{\hat{\alpha}}^{\nu}_{\perp}
\right)\right] 
 \,,  \nonumber\\                            
{\cal L}_{\rm HLS}^{(4)\chi} & = & 
4B \,  w_1 {\rm tr}\left[\hat{\alpha}_{\mu\perp} \hat{\alpha}_\perp^\mu \chi_S \right] 
  + 
 4B \, w_3 {\rm tr}\left[\hat{\alpha}_{\mu \parallel} \hat{\alpha}_{\parallel}^\mu \chi_S \right] + 
  4iB \,  w_5 {\rm tr}\left[(\hat{\alpha}_{\mu \parallel} \hat{\alpha}_\perp^\mu - \hat{\alpha}_{\mu \perp} \hat{\alpha}_{\parallel}^\mu) \chi_P \right] \nonumber\\
& & {} 
+ 
16 B^2 \, w_6 {\rm tr}\left[ \chi_S^2 \right] 
- 
16B^2\, w_8 {\rm tr}\left[ \chi_P^2 \right] 
\,, 
\end{eqnarray} 
\end{widetext} 
where
\begin{eqnarray} 
V_{\mu\nu} & = & 
\partial_\mu V_\nu - \partial_\nu V_\mu - i[V_\mu, V_\nu]
\,, \nonumber \\ 
\hat{\cal V}_{\mu\nu} & = & \frac{1}{2} \left( 
\xi_R^{gg} {\cal R}_{\mu\nu} \xi^{gg \dag}_R + \xi_L^{gg} {\cal L}_{\mu\nu} \xi_L^{gg\dag} \right)  
\,, \nonumber\\
\hat{\cal A}_{\mu\nu} & = &  \frac{1}{2} \left( 
\xi_R^{gg} {\cal R}_{\mu\nu} \xi^{gg\dag}_R - \xi_L^{gg} {\cal L}_{\mu\nu} \xi_L^{gg\dag} 
\right)
\,, \nonumber \\ 
{\cal R}_{\mu\nu} & = & \partial_\mu {\cal R}_\nu - \partial_\nu {\cal R}_\mu - i \left[{\cal R}_\mu, {\cal R}_\nu \right] \,, \nonumber\\ 
{\cal L}_{\mu\nu} & = & \partial_\mu {\cal L}_\nu - \partial_\nu {\cal L}_\mu - i \left[{\cal L}_\mu, {\cal L}_\nu\right] \,, \nonumber\\ 
\hat{\alpha}_{\mu \parallel, \perp} & = & \frac{ 1}{2i} \left( D_\mu \xi_R^{gg} \cdot \xi_R^{gg\dag} \pm D_\mu \xi_L^{gg} \cdot \xi_L^{gg\dag} \right) \,, \nonumber \\ 
D_\mu \xi_{R}^{gg} & = & \partial_\mu \xi_{R}^{gg} - i V_\mu \xi_{R}^{gg} + i \xi_{R}^{gg} {\cal R}_\mu \,,
\nonumber \\ 
D_\mu \xi_{L}^{gg} & = & \partial_\mu \xi_{L}^{gg} - i V_\mu \xi_{L}^{gg} + i \xi_{L}^{gg} {\cal L}_\mu \,. 
\end{eqnarray}

In order to perform the integrating-out procedure, we make the chiral order counting rule in the derivative expansion as follows:
\begin{eqnarray} 
\partial_\mu &\sim& V_\mu^{gg} \sim A_\mu^{gg} \sim {\cal O}(p) 
\,, \nonumber \\ 
S^{gg} &\sim& P^{gg} \sim {\cal O}(p^2)  
\,,
\label{rule}
\end{eqnarray}
which is the same rule as that in the HLS-ChPT/ChPT~\cite{Tanabashi,Harada:2003jx}.  
We identify the lowest KK vector field $V_\mu^{gg(1)}$ (renamed $V_\mu$)  as the lowest-lying vector mesons, $\rho$ and its flavor partners. 
Using the KK decomposition formulas in Eqs.~\eqref{KK:decompose} and \eqref{Vmu:KK:decompose}, we thus solve the equations of motion for the heavier vector and axial-vector fields $V_\mu^{gg(n\ge 2)}$ and $A_\mu^{gg(n \ge 1)}$ 
and expand them in powers of derivatives as 
\begin{eqnarray} 
V_\mu^{gg(n\ge 2)} &=& \alpha^{gg}_{\mu \parallel} + \frac{{\cal O}(p^3)}{m_{V_n^{gg}}^2} 
\,, \nonumber\\ 
A_\mu^{gg(n \ge1)} &=&  \frac{{\cal O}(p^3)}{m_{A_n^{gg}}^2} 
\,. \label{V:A:sol}
\end{eqnarray}
From these we may write the five-dimensional gauge fields $V_\mu^{gg}(x,z)$ and $A_\mu^{gg}(x,z)$ in Eq.~\eqref{KK:decompose} in the form 
\begin{eqnarray} 
V_\mu^{gg}(x,z) & = & 
\alpha^{gg}_{\mu \parallel} + \hat{\alpha}_{\mu \parallel} \psi_1^V + \frac{{\cal O}(p^3)}{m_{V_n^{gg}}^2} 
\,, \nonumber \\ 
A_\mu^{gg}(x,z) & = & 
\hat{\alpha}_{\mu \perp}  \psi_0^A + \frac{{\cal O}(p^3)}{m_{A_n^{gg}}^2 }
\,. 
\end{eqnarray}

For the bulk scalar $\Phi_S^{gg}$, we may introduce an IR potential 
(which will not be specified here) to realize its nonzero VEV of $\Phi_S^{gg}$ 
even in the chiral limit~\cite{Da Rold:2005zs} and allow the scalar and pseudoscalar KK modes 
$S_n^{gg}(x)$ and $P_n^{gg}(x)$ to get the stabilized masses $m_{S_n^{gg}}$ and $m_{P_n^{gg}}$. 
Following the order counting rule in Eq.(\ref{rule}), we thus expand the bulk scalar and pseudoscalar sectors as
\begin{widetext}
\begin{eqnarray} 
 S_{S_n} & = & \frac{1}{g_5^2} \int d^4x dz \, \sum_n 
  {\rm tr} \left[ 
 \left\{  
\left[ 8 a^3 v_S (\psi_0^A)^2 \psi_n^S \right] \hat{\alpha}_{\mu \perp}^2  
- \left[  2 a^5 u_S m_{\Phi_S}^2 \psi_n^S 
+ 2 \dot{u}_S 
a^3 \dot{\psi}_n^S  \right] 
\chi_S \right\} S_n^{gg} 
\right] \nonumber\\
& & {} - \int d^4x \sum_n m_{S_n}^2 {\rm tr}\left[(S_n^{gg})^2\right] + {\cal O}(p^6) 
\,, \nonumber \\ 
 S_{P_n} & = & \frac{1}{g_5^2} \int d^4x dz \, \sum_n {\rm tr} \Bigg[ 
 \Bigg\{  
\left( -4 a^3 v_S \psi_0^A\psi_n^P \right) D_\mu \hat{\alpha}^\mu_\perp 
+\left( 4 a^3 v_S  (1+\psi_1^V)  \psi_0^A  \psi_n^P \right)  i \left[\hat{\alpha}_{\mu \parallel}, \hat{\alpha}_\perp^\mu\right] \nonumber\\
& & \qquad\qquad\qquad\qquad\quad {} -  
\left[
2 a^5 u_S m_{\Phi}^2  
+ 2 \dot{u}_S
a^3 \dot{\psi}_n^P  
\right] \chi_P
\Bigg\} P_n^{gg} \Bigg] \nonumber\\
& & {} - \int d^4x \sum_n m_{P_n}^2 {\rm tr}\left[(P_n^{gg})^2 \right] 
+ {\cal O}(p^6) 
\,,\label{S:P:action}
\end{eqnarray}
where 
\begin{equation} 
 D^\mu \hat{\alpha}_{\mu \perp} 
 \equiv \partial^\mu \hat{\alpha}_{\mu\perp} - i \left[V^\mu, \hat{\alpha}_{\mu\perp}\right]
 \,. 
\end{equation}
Here, the IR boundary terms for $S_n^{gg}$ and $P_n^{gg}$ have been fully eliminated by imposing some IR BCs (as done in~\cite{DaRold:2005vr}),  
which are not specified here. 
From the action in Eq.(\ref{S:P:action}) we derive the equations of motion for $S_n^{gg}$ and $P_n^{gg}$ to find the solutions
\begin{eqnarray} 
 (S_{(n\ge 1)}^{gg})^A & = & \frac{1}{g_5^2} \left[ 
\frac{ {\rm tr}\left[\left(b_n^{s1} \, \hat{\alpha}_{\mu \perp}^2 + b_n^{s2} \, \chi_S \right) T^A\right]}{m_{S_n^{gg}}^2} 
+ \frac{{\cal O}(p^4)}{m_{S_n^{gg}}^2} \right] , \nonumber \\ 
 (P_{(n\ge 1)}^{gg})^A & = & \frac{1}{g_5^2} \left[ 
\frac{ {\rm tr}\left[\left(b_n^{p1} i \left[ \hat{\alpha_{\mu \parallel}}, \hat{\alpha}_{\perp}^\mu \right] + b_n^{p2} \, \chi_P \right) T^A\right]}{m_{P_n^{gg}}^2} + \frac{{\cal O}(p^4)}{m_{P_n^{gg}}^2} 
\right], \label{S:P}
\end{eqnarray}
with  
\begin{eqnarray} 
 b_n^{s1} & = & \left\langle 8 a^2 v_S (\psi_0^A)^2 \psi_n^S \right\rangle  
 \,, \nonumber \\ 
 b_n^{s2} & = & \left \langle - 2 \dot{u}_S 
a^2 \dot{\psi}_n^S 
 - 2 m_{\Phi_S}^2 a^4 
u_S \psi_n^S \right\rangle 
\,, \nonumber \\ 
 b_n^{p1} & = & \left\langle 4 a^2 v_S \left(\frac{F_\sigma^2}{F_\pi^2} + \psi_1^V \right)  \psi_0^A \psi_n^P \right\rangle  
 \,,\nonumber \\ 
 b_n^{p2} &=& \left\langle  
- 2 \dot{u}_S 
a^2 \dot{\psi}_n^P   
- 2 m_{\Phi_S}^2 a^4 
u_S \psi_n^P 
- 4 B a^2 v_S \psi_0^A \psi_n^P \right\rangle 
 \,,  
\end{eqnarray}
\end{widetext}
where $ \dot{A}(z) \equiv \partial_z A(z) $ and 
\begin{eqnarray}   
 \langle A \rangle &\equiv& \int_{z_{\rm UV}}^{z_{\rm IR}} dz a(z) A(z)
\,, 
\end{eqnarray}
 for an arbitrary function $A(z)$. 
In arriving at Eq.(\ref{S:P}) we used 
the equation of motion for $\hat{\alpha}_{\mu \perp}$ derived from the generic HLS Lagrangian Eq.~\eqref{HLS:L2:L4},
\begin{eqnarray} 
 D^\mu \hat{\alpha}_{\mu \perp} 
 & = & {} - i \left( \frac{F_\sigma^2}{F_\pi^2} - 1 \right) \left[\hat{\alpha}_{\mu \parallel}, \hat{\alpha}^{\mu}_\perp\right] 
+ B \chi_P + {\cal O}(p^4) 
 \,. \nonumber\\
\end{eqnarray}

By putting the solutions in Eqs.~\eqref{V:A:sol} and \eqref{S:P} back into the DBI action, Eq.~\eqref{DBI:action2}, we thus find 
\begin{eqnarray}
S_{\rm DBI}
= \int d^4x \left({\cal L}_{\rm HLS} + {\cal O}(p^6)\right) 
+ S[v_S] 
\,, 
\end{eqnarray}
where 
\begin{eqnarray}
 S[v_S] 
 = {} - \frac{1}{g_5^2} \int d^4 x \int_{z_{\rm UV}}^{z_{\rm IR}} dz \, 
 a^3 \left[ \left(\partial_z v_S\right)^2 + a^2 m_{\Phi_S}^2 v_S^2 \right] 
\,.  \nonumber\\
\label{S:Vs}
\end{eqnarray}
Consequently, the coefficients of terms in the general HLS Lagrangian ${\cal L}_{\rm HLS}$ given in Eq.~\eqref{HLS:L2:L4} are now 
expressed by the hQCD model quantities written in terms of the integrals over the $z$ direction (for convenience and simplicity, we have redefined $\psi_1^V$ as $-\psi_1^V$),\footnote{
 Using the equation of motion for $\psi_0^A$ in Eq.~\eqref{EOM:psi0A}, 
with the BC, $\psi_0^A(z_{\rm UV})=1$ and $\dot{\psi}_0^A(z=z_{\rm IR})=0$, 
we may rewrite $F_\pi$ in Eq.~\eqref{eq:lecshqcd} as 
$ 
F_\pi^2 
 = - \frac{1}{g_5^2} a(z_{\rm UV}) \dot{ \psi}_0^A(z=z_{\rm UV}) 
$, which is in agreement with the chiral-limit expression obtained in Ref.~\cite{Erlich:2005qh} with $a(z)=1/z$. 
}
\begin{widetext}
\begin{eqnarray} 
F_{\pi}^2 & = & \frac{1}{g_5^2} \left\langle \left[ \dot{\psi}_0^A\right]^2 + 4 a^2 v_S^2 \left[\psi_0^A\right]^2 \right\rangle 
\,, \quad F_{\sigma}^2 = \frac{1}{g_5^2} m_\rho^2 \left\langle \left[\psi^V_1\right]^2 \right\rangle 
\, , \quad \frac{1}{g^2} = \frac{1}{g_5^2}  \left\langle \left[\psi_1^V\right]^2 \right\rangle 
\,, \nonumber \\
B F_\pi^2 & = & \frac{2}{g_5^2} \left\langle  a^2 \left( 
- \dot{u}_S \partial_z 
- a^2  m_{\Phi_S}^2 u_S \right) 
v_S \right\rangle 
\,, \nonumber\\
y_1 & = & {} -\frac{1}{g_5^2} \left\langle (1 + \psi_1^V - [\psi_0^A]^2)^2 \right\rangle 
+ \frac{1}{4g_5^4} \sum_n \frac{\left[b_n^{s1}\right]^2}{m_{S_n^{gg}}^2} 
\,, \quad y_2 = \frac{1}{g_5^2} \left\langle (1 + \psi_1^V - [\psi_0^A]^2)^2 \right\rangle 
\,, \nonumber\\ 
y_3 & = & {} - y_4 = {} - \frac{1}{g_5^2} \left\langle \left[\psi^V_1\right]^2 \left(1 + \psi_1^V \right)^2 \right\rangle 
\,, \quad y_5 = {} - \frac{2}{g_5^2} \left\langle \left[\psi_1^V\right]^2 \left[\psi_0^A\right]^2 \right\rangle 
\, , \nonumber \\
y_6 & = & \frac{2}{g_5^2} \left\langle  \left[\psi_1^V\right]^2 \left[\psi_0^A\right]^2 -  \psi_1^V \left ( 1 + \psi_1^V \right) \left(1 + \psi_1^V - \left[\psi_0^A\right]^2 \right)\right\rangle 
\,, \nonumber \\
y_7 & = & \frac{2}{g_5^2} \left\langle \psi_1^V \left (1 + \psi_1^V\right)\left(1 + \psi_1^V - \left[\psi_0^A\right]^2 \right) \right\rangle 
+ \frac{1}{2 g_5^4} \sum_n \frac{\left[b_n^{p1}\right]^2}{m_{P_n^{gg}}^2}
\,, \nonumber \\ 
y_8 & = & {} - \frac{1}{g_5^2} \left\langle \left[\psi_1^V\right]^2 \left[\psi_0^A\right]^2 \right\rangle 
- \frac{1}{4 g_5^4} \sum_n \frac{\left[b_n^{p1}\right]^2}{m_{P_n^{gg}}^2}
\,, \quad y_9 = \frac{2}{g_5^2} \left\langle \left[\psi_1^V\right]^2 \left[\psi_0^A\right]^2 \right\rangle
\,, \nonumber\\
z_1 & = & {} - \frac{1}{2} \frac{1}{g_5^2}  
\left\langle ( 1+\psi_1^V)^2 \right\rangle 
\,, \quad z_2 = {} - \frac{1}{2} \frac{1}{g_5^2} \left\langle \left[\psi_0^A\right]^2 \right\rangle 
\,, \quad z_3 = \frac{1}{g_5^2} \left\langle {\psi}_1^V \left( 1+ {\psi}_1^V\right) \right\rangle 
\,, \nonumber\\
z_4 & = & 2 \frac{1}{g_5^2} \left\langle {\psi}_1^V \left(1+{\psi}_1^V - \left[\psi_0^A\right]^2 \right) \right\rangle 
\, , \quad z_5 = {} - 2 \frac{1}{g_5^2} \left\langle \left[{\psi}_1^V\right]^2 \left( 1+{\psi}_1^V \right) \right\rangle 
\,, \nonumber\\
z_6 & = & {} -2 \frac{1}{g_5^2} \left\langle \left( 1+{\psi}_1^V - \left[\psi_0^A\right]^2 \right)\left( 1+{\psi}_1^V \right)\right\rangle 
\,,\quad z_7 = 2 \frac{1}{g_5^2} \left\langle {\psi}_1^V \left( 1+{\psi}_1^V \right)^2 \right \rangle 
\,, \quad 
z_8 ={} -2 \frac{1}{g_5^2} \left\langle {\psi}_1^V \left[\psi_0^A\right]^2\right\rangle 
\,, \nonumber\\
4 B \, w_1 & = & 
\frac{8}{g_5^2} \left\langle a^2 v_S 
u_S 
\left[\psi_0^A\right]^2 \right\rangle
+ \frac{1}{2 g_5^4} \sum_n \frac{b_n^{s1}b_n^{s2}}{m_{S_n^{gg}}^2} 
\,, \quad 
4 B \, w_3 =  0 
\,, \nonumber\\ 
4 B \, w_5 & = & \frac{4}{g_5^2} \left\langle 
a^2 v_S 
u_S 
\psi_0^A  \left[ \frac{F_\sigma^2}{F_\pi^2} 
+ \psi_1^V \right] \right\rangle
+ \frac{1}{2 g_5^4} \sum_n \frac{b_n^{p1} b_n^{p2}}{m_{P_n^{gg}}^2}
\,, \nonumber\\ 
16 B^2\, w_6 & = & {} - \frac{1}{g_5^2} \left\langle a^2 \left[ 
\dot{u}_S^2 + a^2 u_S^2 
m_{\Phi_S}^2 \right] \right\rangle 
+ \frac{1}{4 g_5^4} \sum_n \frac{\left[b_n^{s2} \right]^2}{m_{S_n^{gg}}^2}
\,, \nonumber\\ 
16 B^2 \, w_8  
& = & 
\frac{1}{g_5^2}  4 B \left\langle  a^2 v_S 
u_S 
\psi_0^A  \right\rangle 
+ \frac{1}{g_5^2} \left\langle a^2 \left[ \dot{u}_S^2 
+ a^2 
u_S^2 
m_{\Phi_S}^2 \right] \right\rangle 
 - \frac{1}{4 g_5^4} \sum_n \frac{\left[b_n^{p2}\right]^2}{m_{P_n^{gg}}^2}
\,. 
\label{eq:lecshqcd}
\end{eqnarray} 
\end{widetext}
Note the absence of the $w_3$ term because we have assumed flavor symmetry, i.e, $v_S \propto  \mathbf{1}_{N_f \times N_f}$, 
in which case there is no coupling between the bulk vector field and the scalar VEV in the original action Eq.~\eqref{DBI:action2}. 
If the flavor symmetry breaking effect was taking into account, we would have nonzero $w_3$.

Using the expressions of the low-energy constants of HLS given in Eq.~\eqref{eq:lecshqcd}, one can compute the 
numerical values which are useful for phenomenological applications. 
To do this, we need to specify the IR potential of $\Phi_S$,  
for example, like the one proposed in Ref.~\cite{DaRold:2005vr}. 
We leave such a concrete model calculation for future work 
since the present subject is to propose a procedure to obtain the chiral effective theories from hQCD models.

The normalizable solution of the VEV, $v_S(z)$, can be obtained by solving Eq.(\ref{PhiS0}) with $\Phi_S^0(z)=v_S(z)$ along with the normalizable BC,
\begin{eqnarray} 
\frac{v_S(z)}{u_S(z)} \Bigg|_{z=z_{\rm UV}} 
& = & 0
\,, \quad v_S(z) |_{z=z_{\rm IR}} = \frac{\xi}{L} 
\,,   
\label{eq:BCsvevsn}
\end{eqnarray}
where $\xi$ denotes the IR value of $v_S(z)$ which is related to the chiral condensate $\langle \bar{q}q \rangle$. 
(We take $\xi$ to be negative so as to have the negative chiral condensate $\langle \bar{q}q \rangle <0$.)  
Note that the UV BC in Eq.~\eqref{eq:BCsvevsn} 
follows from the parametrization of $\Phi_S$ in Eq.~\eqref{eq:paraSgg} with the BC in Eq.~\eqref{eq:BCSP}.

To be specific, we take the standard AdS space, i.e., $a(z) = L/z$, to explicitly solve the equation of motion,
Eq.(\ref{PhiS0}). 
The explicit expression of the function $u_S(z)$ is then given as 
the non-normalizable solution of Eq.(\ref{PhiS0}), which turns out to be 
$u_S(z) = (z/L)^{4-\Delta}$. 
This leads to 
\begin{eqnarray} 
 v_S(z) &=& 
 \frac{\xi}{L} 
\left( \frac{z}{z_{\rm IR}}  \right)^\Delta 
\frac{1-(z_{\rm UV}/z)^{2(\Delta -2)}}{1-(z_{\rm UV}/z_{\rm IR})^{2(\Delta -2)}} 
\,. 
\label{VEV:sol}
\end{eqnarray}   
Substituting the solution given by Eq.~(\ref{VEV:sol}) into the $(B F_\pi^2)$ term in Eq.~\eqref{eq:lecshqcd}, we obtain
\begin{eqnarray} 
 BF_\pi^2 = {}
- \frac{L}{g_5^2} \frac{2 (4 - \Delta) \xi}{z_{\rm IR}^3} \left( \frac{z_{\rm IR}}{L} \right)^{3-\Delta}
\,. 
\end{eqnarray} 
On the other hand, by using the functional derivative, we may identify the chiral condensate $\langle \bar{q}q \rangle$ as 
\begin{eqnarray}
\frac{\delta S_{\rm DBI}}{\delta {\cal M}} \Bigg|_{{\cal M}=0} = {} - \langle \bar{q}q \rangle  
\,.
\end{eqnarray}
The holographic correspondence between the QCD generating functional and the hQCD action then yields 
\begin{equation} 
  \langle \bar{q} q \rangle = - B F_\pi^2 = \frac{L}{g_5^2} \frac{2 (4 - \Delta) \xi}{z_{\rm IR}^3} \left( \frac{z_{\rm IR}}{L} \right)^{3-\Delta}
  \,. \label{xi:chiralcond}
\end{equation}
Consequently, the $BF_\pi^2$ term in ${\cal L}_{\rm HLS}^{(2)\chi}$ is reexpressed as
\begin{equation} 
{} - \langle \bar{q}q  \rangle {\rm tr}[\chi_S]  \,.  
\end{equation} 
Taking $N_f = 2$ and expanding the $\xi_{L,R}^{gg}$ fields in $\chi_S$, we thus see that 
at the $O(p^2)$ level the GOR relation is manifestly realized as
\begin{equation} 
  m_\pi^2  F_\pi^2 = - (m_u+m_d) \langle \bar{q}q \rangle 
  \,, 
\end{equation}
in agreement with Refs.~\cite{Erlich:2005qh,DaRold:2005vr}. 
The GOR relation, of course, gets corrections from terms of
 $\mathcal{O}(p^4)$~\cite{Nishihara:2014nva} in a proper manner, as in the ChPT.

If one worked on a metric other than the AdS, 
a similar discussion could be done once the warping factor $a(z)$, the associated  
the non-normalizable solution $u_S(z)$, and the normalizable one $v_S(z)$ are specified.

\section{The IP-odd sector of the hidden local symmetry Lagrangian} 

\label{sec:cshls}

We next move on to the CS term in Eq.~\eqref{eq:CSpart} and attempt to integrate out the heavy KK modes to obtain the IP-odd sector of the HLS Lagrangian~\cite{Fujiwara:1984mp,Harada:2003jx}.

Before making the explicit integrating-out procedure, 
we first consider the gauged-chiral transformation of the CS part in Eq.~\eqref{eq:CSpart} 
with the infinitesimal parameters $\Lambda_{L,R}$: 
\begin{eqnarray}
\delta_{\Lambda_{L,R}} S_{\rm CS} 
& = & \frac{N_c}{24 \pi^2}  \int_{M^4}  \int_{z_{\rm UV}}^{z_{\rm IR}} 
d \left[ \omega_4(\Lambda_R, R) - \omega_4 (\Lambda_L, L)  \right] \nonumber \\ 
& = & \int_{M^4} \Big\{ \big[ \omega_4(\Lambda_L, L) - \omega_4(\Lambda_R,R) \big]_{z=z_{\rm UV}}
\nonumber \\ 
& & \qquad\;\; {} - \big[ \omega_4(\Lambda_L, L) - \omega_4(\Lambda_R, R) \big]_{z=z_{\rm IR}}
\Big\}\,,  \nonumber\\
\end{eqnarray}
where the first term exactly reproduces the desired chiral anomaly in QCD,  
while the  second term gives an extra contribution. 
To cancel it, one needs to add a counterterm  $S_{\rm counter}$ which satisfies 
\begin{equation} 
\delta_{\Lambda_{L,R}} S_{\rm counter} 
= {} \int_{M^4} 
 \Big\{ \omega_4(\Lambda_L, L) - \omega_4(\Lambda_R,R) \Big\}_{z=z_{\rm IR}} 
\,. 
\end{equation}

We next explicitly determine the counterterm in the $L_z = R_z = 0$ gauge. In this gauge the CS term takes the following form:
\begin{eqnarray} 
S_{\rm CS} & = & S_{\rm CS}^{(1)} + S_{\rm CS}^{(2)} + S_{\rm CS}^{(3)}
\,, \label{CS:gauge:fixed}
\end{eqnarray}
 where
\begin{eqnarray} 
S_{\rm CS}^{(1)}
& = &{} -
\frac{N_c}{240\pi^2} \int_{M^4 \times z} \left[ {\rm tr}\left(ig_R^g d g_R^{g\dag}\right)^5 - (R \leftrightarrow L) \right], \nonumber\\
S_{\rm CS}^{(2)}
& = &{} -
\frac{N_c}{24\pi^2} \int_{M^4 \times z} \left[  d \alpha_4\left(i dg_R^{g\dag} g_R^g, R\right) - (R \leftrightarrow L) \right], \nonumber\\
S_{\rm CS}^{(3)}
& = &
\frac{N_c}{24\pi^2} \int_{M^4 \times z} \left[ \omega_5\left(R^g\right) -  (R \leftrightarrow L)  \right]
\,, \label{CS:gauge:fixedSi}
\end{eqnarray}
with the definitions
\begin{eqnarray} 
\omega_5(A^g) & = & {\rm tr}\left[ A^g d A^g d A^g - \frac{3}{2} i (A^g)^3  d A^g\right] 
\,, \nonumber \\ 
\alpha_4(V,A) & = &{} - \frac{1}{2}{\rm tr}\Big[V \left(iAdA + idAA + A^3\right) \nonumber\\
& & \qquad\quad{} - \frac{1}{2} VAVA - V^3 A\Big] 
\,.
\end{eqnarray}
The definitions of $g_R^g$ and $g_L^g$ in Eq.~\eqref{CS:gauge:fixedSi} are given in Eq.~\eqref{eq:gaugez0}.

The first term in Eq.~\eqref{CS:gauge:fixed} can be rewritten into a total derivative independent of gauge configuration. Since $g_{R,L}^g(z=z_{\rm UV})=\xi_{R,L}^g$ and $g_{R,L}^g(z=z_{\rm IR})=1$, $S_{\rm CS}^{(1)}$ is only written in terms of $\xi_{R,L}^g$. 
Fixing the local-chiral gauge by taking $\xi_R^g=1$ and $\xi_L^g=U^\dag$, we thus find that the $S_{\rm CS}^{(1)}$ term is identical to the Wess-Zumino term
\begin{eqnarray}
S_{\rm CS}^{(1)} & = & {}
- \frac{N_c}{240\pi^2} \int_{M^5} {\rm tr}[i dUU^\dag]^5 
= \Gamma_{\rm WZ}[U] 
\,. \label{WZ:term}
\end{eqnarray}  
The $S_{\rm CS}^{(2)} $ term in Eq.~\eqref{CS:gauge:fixed} is evaluated as 
\begin{eqnarray} 
S_{\rm CS}^{(2)} & = & 
 \frac{N_c}{24\pi^2} \int_{M^4} 
 \left[ \alpha_4(i dg_L^{g\dag}\cdot g_L^g, L) -    (R \leftrightarrow L) \right]^{z=z_{\rm IR}}_{z=z_{\rm UV}} 
\nonumber \\ 
& = & 
\frac{N_c}{24\pi^2} \int_{M^4 \times z} 
\left[ \alpha_4(i d\xi_R^{g\dag} \cdot\xi_R^g, R) -    (R \leftrightarrow L) \right]_{z=z_{\rm UV}} . \nonumber\\
\label{alpha4:term}
\end{eqnarray}

Combining all three terms in~\eqref{CS:gauge:fixed}, one thus finds the CS term takes the form
\begin{equation} 
S_{\rm CS} = \Gamma_{\rm WZW} + \Gamma_{\rm HLS}^{\rm inv} + \Gamma_{\rm IR} 
\,, \label{eq:CS4dtotal}
\end{equation}
where $\Gamma_{\rm WZW} $ stands for the (covariantized) Wess-Zumino-Witten term~\cite{Wess:1971yu}. 
$\Gamma_{\rm HLS}^{\rm inv}$ denotes the HLS/chiral invariant term~\cite{Harada:2010cn} expressed as 
\begin{eqnarray} 
\Gamma_{\rm HLS}^{\rm inv} 
& = & \frac{N_c}{12 \pi^2} \int_{M^4 \times z} {\rm tr}\Bigg[ 
3 A^g dV^gdV^g + A^g dA^gdA^g 
\nonumber \\ 
&& - 3 i \left((V^g)^2A^g+A^g(V^g)^2+(A^g)^3\right) dV^g 
\nonumber \\ 
&& - 3i (A^gV^gA^gdA^g)_{\rm nonzero} 
\Bigg]  
\,,
\label{HLS:inv:action}
\end{eqnarray}
where ``nonzero" means 
terms including at least one massive vector or axial-vector field. 
The last term, $\Gamma_{\rm IR}$, 
in Eq.~\eqref{eq:CS4dtotal} gives an extra contribution to the $\pi^0$-$\gamma$-$\gamma$ vertex 
which is controlled by the low-energy theorem of chiral dynamics. 
Through explicit calculations, we obtain
\begin{widetext}
\begin{eqnarray} 
\Gamma_{\rm IR} & = & {} - \frac{N_c}{12 \pi^2} \int_{M^4} {\rm tr}
\left[\left(V^gdV^g+dV^gV^g\right) A^g - \frac{3}{4} i \left((V^g)^3 A^g + V^g (A^g)^3\right)\right]^{z=z_{\rm IR}} 
+ \frac{iN_c}{12\pi^2} \int_{M^4} {\rm tr}\left[ \alpha^g_{\parallel} \alpha_\perp^{g3} \right] \left[\psi_0^A(z_{\rm IR})\right]^3 \nonumber\\
& & {} - \frac{N_c}{12\pi^2} \int_{M^4} {\rm tr}\left[(A^g)^3 V^g \right]_{\rm nonzero}^{z=z_{\rm IR}} 
\,. \label{IR:action}
\end{eqnarray}
\end{widetext}
To cancel this term, we introduce a counterterm $\Gamma_c$ in such a way that 
\begin{equation}
\Gamma_{\rm IR} + \Gamma_c = 0 
\,. 
\label{subtract}
\end{equation}

We shall now integrate out the heavier vector and axial-vector meson fields from the CS term, except the lowest-lying vector meson fields, in a way similar to the DBI part.  
For this purpose, all we should do is fix the gauges except for the HLS regarding the lightest vector meson $\rho$ (and its flavor partners) and 
set the five-dimensional vector and axial-vector fields as
$V^{gg}= \alpha_{\parallel}^{gg} +  \hat{\alpha}_{\parallel} \psi_1^V$ and $A^{gg} = \hat{\alpha}_\perp \psi_0^A$ and substitute them into 
the HLS invariant action $\Gamma_{\rm HLS}^{\rm inv}$ in Eq.~\eqref{HLS:inv:action}. Then we have 
\begin{widetext}
\begin{eqnarray} 
\Gamma_{\rm HLS}^{\rm inv} & = & \frac{N_c}{12 \pi^2} \int_{M^4 \times z} {\rm tr} [3 A^{gg} dV^{gg}dV^{gg} + A^{gg} dA^{gg}dA^{gg} 
\nonumber\\ 
&& 
- 3 i \left((V^{gg})^2A^{gg} + A^{gg}(V^{gg})^2+(A^{gg})^3\right) dV^{gg} - 3i (A^{gg}V^{gg}A^{gg}dA^{gg})_{\rm nonzero} 
] 
\nonumber \\ 
& = & {} - \frac{N_c}{4 \pi^2} \int_{M^4} \int_{z_{\rm UV}}^{z_{\rm IR}} dz \, {\rm tr} [ A^{gg} \partial_z V^{gg} dV^{gg} 
+ A^{gg} d V^{gg} \partial_z V^{gg} 
\nonumber \\ 
&& 
- i \left( (V^{gg})^2 A^{gg} + A^{gg} (V^{gg})^2 + (A^{gg})^3 \right) \partial_z V^{gg} - i (A^{gg} V^{gg}A^{gg} \partial_z A^{gg})_{\rm nonzero} 
] 
\,. 
\end{eqnarray} 
We thus find 
\begin{eqnarray} 
\Gamma_{\rm HLS}^{\rm inv} 
& = &  \frac{N_c}{24\pi^2}  \int_{M^4} 
\left\{ 
x_1 {\rm tr}\left[\left( \hat{\alpha}_\perp \hat{\alpha}_{\parallel} - \hat{\alpha}_{\parallel} \hat{\alpha}_{\perp} \right) F_V \right]
+ i x_2 {\rm tr}\left[ \hat{\alpha}_\perp  \hat{\alpha}_{\parallel}^3 \right] + i x_3 {\rm tr} \left[\hat{\alpha}_{\parallel}  \hat{\alpha}_{\perp}^3 \right] + x_4 {\rm tr}\left[\left( \hat{\alpha}_\perp \hat{\alpha}_{\parallel} - \hat{\alpha}_{\parallel} \hat{\alpha}_{\perp} \right) \hat{F}_V \right] \right\} 
\,, \label{gamma3:def}
\end{eqnarray}
\end{widetext}
 where the coefficients $x_i (i=1,2,3,4)$ are expressed in terms of hQCD quantities as 
\begin{eqnarray} 
x_1 &=& \left\langle\left\langle  6 \, \psi_1^V \psi_0^A \dot{\psi}_1^V \right\rangle\right\rangle 
\,, \nonumber\\ 
x_2 &=& \left\langle\left\langle  12 \, \psi_1^V \psi_0^A \dot{\psi}_1^V (1 + \psi_1^V) \right\rangle\right\rangle 
\,, \nonumber\\ 
x_3 &=& \left\langle\left\langle 4 \, \psi_0^A \dot{\psi}_1^V ( - [\psi_0^A]^2 + 3 \psi_1^V + 3) \right\rangle\right\rangle 
\,, \nonumber\\ 
x_4 &=& \left\langle\left\langle  - 6 \, (\psi_1^V + 1) \psi_0^A \dot{\psi}_1^V \right\rangle\right\rangle 
\, ,
\end{eqnarray} 
with the definition
\begin{equation} 
\langle\langle A \rangle\rangle \equiv \int_{z_{\rm UV}}^{z_{\rm IR}} dz A(z) 
\,, 
\end{equation} 
for an arbitrary function $A(z)$. 
The $c_1$-$c_4$ coefficients in the literature~\cite{Harada:2003jx} are expressed in terms of $x_i$ as
\begin{eqnarray} 
c_1 & = & {} - \frac{1}{12} x_2 + \frac{1}{12} x_3 
\,, \nonumber\\ 
c_2 & = & {} - \frac{1}{12} x_2 - \frac{1}{12} x_3 
\,, \nonumber\\ 
c_3 & = & {} - \frac{1}{3} x_1 
\,,\nonumber\\ 
c_4 & = & {} - \frac{1}{3} x_4 
\,.  
\end{eqnarray}  
Actually, the overall coefficients of $x_1,x_2, x_3$ and $x_4$ (or $c_1, c_2, c_3$ and $c_4$)  
are different by a factor 2 from those reported in Ref.~\cite{Harada:2010cn} based on 
the Sakai-Sugimoto model. This reflects the ambiguity in the subtracting scheme in Eq.(\ref{subtract}), as will be argued later.

\section{Chiral perturbation theory from bottom-up hQCD models} 

\label{sec:chpt}

We may further integrate out the vector meson fields to derive  
the ChPT described only by the NGBs. 
Repeating the procedure done above and using the formulas relating the HLS to the ChPT 
given in Ref.~\cite{Harada:2003jx}, we thus find 
\begin{eqnarray} 
{\cal L}_{\rm ChPT} & = & {} {\cal L}_{\rm ChPT}^{(2)} + {\cal L}_{\rm ChPT}^{(4)}
\, ,
\end{eqnarray}
where
\begin{widetext}
\begin{eqnarray} 
{\cal L}_{\rm ChPT}^{(2)} & = & \frac{F_\pi^2}{4}{\rm tr}\left[ D_\mu U D^\mu U^\dag \right] + \frac{BF_\pi^2}{8}{\rm tr} \left(\chi^\dag U + \chi U^\dag\right) \, ,\nonumber\\
{\cal L}_{\rm ChPT}^{(4)} & = & L_1 \left({\rm tr}\left[D_\mu U^\dag D^\mu U\right]\right)^2 + L_2 \left({\rm tr}\left[D_\mu U^\dag D_\nu U\right]\right)^2
+ L_3 {\rm tr}\left[D_\mu U^\dag D^\mu U)^2\right] 
+ 
L_4 {\rm tr}\left[D_\mu U^\dag D^\mu U\right] {\rm tr}\left[\chi^\dag U  + {\rm h.c.}\right] \nonumber \\ 
& & {} + L_5 {\rm tr}\left[D_\mu U^\dag D^\mu U\left(\chi^\dag U + {\rm h.c.}\right)\right] + 
L_6 \left({\rm tr}\left[\chi^\dag U + {\rm h.c.}\right]\right)^2
+ 
L_7 \left({\rm tr}\left[\chi^\dag U - \chi U^\dag\right]\right)^2 
+ 
L_8 {\rm tr}\left[\chi^\dag U \chi^\dag U + {\rm h.c.}\right] 
\nonumber \\ 
& &{} - i L_9 {\rm tr}\left[{\cal L}_{\mu\nu} D^\mu U D^\nu U^\dag + {\cal R}_{\mu\nu} D^\mu U^\dag D^\nu U\right] 
+ 
L_{10} {\rm tr}\left[U^\dag {\cal L}_{\mu\nu} U R^{\mu\nu}\right]  
+ 
H_1 {\rm tr}\left[{\cal L}_{\mu\nu}^2 + {\cal R}_{\mu\nu}^2\right] 
+ 
H_2 {\rm tr}\left[\chi^\dag \chi\right] 
\,, 
\end{eqnarray} 
with $D_\mu U= \partial_\mu U - i {\cal L}_\mu U + i U {\cal R}_\mu$ and $\chi = 2 B {\cal M}$. 
In terms of the hQCD quantities, the coefficients of the $\mathcal{O}(p^2)$ terms are the same as those given in Eq.~\eqref{eq:lecshqcd}, and the $\mathcal{O}(p^4)$ coefficients are
\begin{eqnarray} 
L_1 & = & \frac{1}{32} \left( \frac{1}{g^2} - z_4 + y_2  \right) = \frac{1}{32g_5^2} \left\langle  \left[1- \left(\psi_0^A\right)^2\right]^2 \right\rangle  
\,, \quad 
L_2 = 
\frac{1}{16} \left( \frac{1}{g^2} - z_4 + y_2 \right) = \frac{1}{16g_5^2} \left\langle  \left[1 - \left(\psi_0^A\right)^2\right]^2 \right\rangle  
\,, \nonumber \\ 
L_3 & = & 
\frac{1}{16} \left( - \frac{3}{g^2} + 3 z_4 + y_1 - 2 y_2 \right)
={} - \frac{3}{16g_5^2} \left\langle \left[1 - \left(\psi_0^A\right)^2\right]^2 \right\rangle + \frac{1}{64 g_5^4} \sum_n \frac{\left(b_n^{s_1}\right)^2}{m_{S_n^{gg}}^2}
\,, \nonumber \\ 
L_4 & = & \frac{1}{4} w_2 
= 0  
\,, \quad
L_5 = \frac{1}{4} w_1 =
\frac{1}{B} \left[ \frac{1}{2g_5^2} \left\langle  a^2 v_S u_S \left( \psi_0^A\right)^2 \right\rangle  
+ \frac{1}{32g_5^4} \sum_n \frac{b_n^{s_1} b_n^{s_2}}{m_{S_n^{gg}}^2} \right] 
\,, \nonumber \\ 
L_6 & = & w_7 
=0 \, ,\quad 
L_7 = w_9 
= 0 
\,, \nonumber \\ 
L_8 & = & w_6 + w_8 =\frac{1}{B} \left[ \frac{1}{g_5^2} \left\langle a^2 v_S u_S \psi_0^A 
\right\rangle \right] 
+ \frac{1}{64B^2g_5^4} \sum_n \left[ \frac{\left(b_n^{s2}\right)^2}{m_{S_n^{gg}}^2} - \frac{\left(b_n^{p2}\right)^2}{m_{P_n^{gg}}^2} \right]  
\,, \nonumber \\ 
L_9 & = & \frac{1}{8} \left( \frac{2}{g^2} - 2 z_3 - z_4 -z_6 \right)
= \frac{1}{4g_5^2} \left\langle 1 - \left(\psi_0^A\right)^2 \right\rangle 
\,,  \quad
L_{10} =
\frac{1}{4} \left( - \frac{1}{g^2} + 2z_3 - 2z_2 + 2z_1 \right)
={} - \frac{1}{4g_5^2} \left\langle 1 - \left(\psi_0^A\right)^2 \right\rangle , \nonumber \\ 
H_1 & = & 
\frac{1}{8} \left( - \frac{1}{g^2} + 2 z_3 + 2 z_2 + 2z_1 \right) 
={} - \frac{1}{8g_5^2} \left\langle 1 + \left(\psi_0^A\right)^2 \right\rangle
\,, \nonumber \\ 
H_2 & = &  2(w_6 -w_8) =
\frac{1}{B}\left[- \frac{2}{g_5^2} \left\langle  a^2 v_S u_S \psi_0^A 
\right\rangle \right] \nonumber\\
& & \qquad\qquad\quad\;\; {} + \frac{1}{B^2} \left\{ \frac{1}{4g_5^2} \left\langle a^2 \left[ 
\dot{u}_S^2 + a^2 u_S^2 m_{\Phi_S}^2 
\right] \right\rangle  
 + 
\frac{1}{32 g_5^4} \sum_n \left( \frac{\left(b_n^{s2}\right)^2}{m_{S_n^{gg}}^2} + \frac{\left(b_n^{p2}\right)^2}{m_{P_n^{gg}}^2} \right) \right\}, 
\label{eq:lecschpt}
 \end{eqnarray}
where the parameter $B$ is determined by Eq.~(\ref{xi:chiralcond}). 
\end{widetext}
As previously mentioned, 
all the low-energy constants given in Eq.~\eqref{eq:lecschpt} can numerically be computed 
once an explicit  expression of the bulk scalar potential is specified. 
However, even without specifying the bulk scalar potential, 
we can draw several interesting observations through our derivations: 

\begin{itemize}

\item The coefficients $L_i$ satisfy relations $2L_1 - L_2 = L_4 = L_6 = 0$ and $L_7=0$ which are consistent with the large $N_c$ limit without the $\eta'$-exchange contribution to $L_7$~\cite{Gasser:1983yg}.

\item The $\psi_1^V$ dependence in $L_i$ completely disappears in our integrating-out procedure. 
This can be understood by noting that our integrating-out procedure 
is equivalent to eliminating the massive KK modes, 
i.e., setting $\psi_n^{V} = \psi_n^{A} = 0 $ in Eq.~\eqref{KK:decompose}.

\item  A holographic prediction $L_9 + L_{10} = 0$ has been derived, which allows us to derive a new sum rule 
\begin{eqnarray}
 \sum_{n \ge 1} \left[ 
 \frac{F_{A_n}^2}{m^2_{A_n}}
 - 
 \frac{F_{V_n}^2}{m^2_{V_n}} 
 + 
 \frac{2 F_\pi^2 F_{V_n} g_{V_n \pi\pi}}{ m_{V_n}^3} 
\right] 
= 0 
\,, 
\end{eqnarray}
where we make use of the following infinite-sum expressions for $L_{9}$ [i.e., charge radius of pion $\langle r^2_{\pi^\pm} \rangle$] and $L_{10}$ 
(i.e., difference of vector and axial-vector current correlators $\Pi_{V-A}$), 
\begin{eqnarray}
L_9  & = & \frac{F_\pi^2}{12} \langle r^2_{\pi^\pm} \rangle 
= \frac{F_\pi^2}{2} \sum_{n\ge 1} \frac{F_{V_n} g_{V_n \pi\pi}}{m_{V_n}^3}
\,, \nonumber \\ 
L_{10} & = & \frac{1}{4} \frac{d}{d Q^2} \left[ \Pi_A(Q^2) - \Pi_V(Q^2)  \right] \Bigg|_{Q^2=0} \nonumber\\
& = & \frac{1}{4} \sum_{n\ge 1} \left[ 
 \frac{F_{A_n}^2}{m_{A_n}^2} - \frac{F_{V_n}^2}{m_{V_n}^2} 
\right] 
\,, 
\end{eqnarray}
with the $V_n$ and $A_n$ decay widths $F_{V_n}$ and $F_{A_n}$, and the $V_n$-$\pi$-$\pi$ coupling constant $g_{V_n \pi\pi}$. 
(A similar sum rule was derived in Ref.~\cite{Hirn:2005nr} from a hQCD model without bulk scalars. )
 The holographic prediction $(L_9 + L_{10})=0$ gives a phenomenological implication: 
The pion axial-vector form factor $F_A$ vanishes,  
which is, however, disfavored by the experiments~\cite{Beringer:1900zz}. 
This point was first argued in Ref.~\cite{Da Rold:2005zs}, on a holographic setup similar to the present hQCD with bulk scalars, 
which was only through a numerical estimate for $L_9$ and $L_{10}$.

\end{itemize}

\section{Summary}

\label{sec:dis}

 In this work, we proposed a gauge-invariant method to integrate out the heavy KK modes from a class of bottom-up 
hQCD models including bulk scalars. We integrated out heavy KK modes from the hQCD model 
to arrive at the chiral effective theory, HLS-ChPT, based on 
the derivative expansion, including the NGBs and the lowest-lying vector mesons $\rho$ and their flavor partners. 
All the coefficients of the chiral effective theories were then expressed in terms of a few intrinsic quantities of hQCD models. 
The GOR relation was realized properly at the ${\cal O}(p^2)$ level of the HLS-ChPT reduced from the hQCD model.

We also found that, to reproduce the desired chiral anomaly in QCD, not only the CS term but also some counterterms should be included 
because of the nontrivial IR values of the five-dimensional gauge fields. The expressions of the counterterms were explicitly presented in terms of the five-dimensional gauge fields [Eq.~(\ref{subtract})].

The counterterm $\Gamma_{\rm c}$ introduced in Eq.(\ref{subtract}) is actually not unique in canceling the extra contribution from 
the IR term $\Gamma_{\rm IR}$: It can involve terms invariant under the hidden local/chiral symmetry. 
This implies that the coefficients $c_1, c_2, c_3$ and $c_4$ in the HLS-invariant action $\Gamma_{\rm HLS}$ are actually 
not determined from the present setup of the hQCD model. 
Instead, one needs some phenomenological (at least four) inputs 
to discuss the IP-odd processes, including vector mesons, from the model. 
Once some phenomenological inputs are set, however, one could make some IP-odd vertices,  
calculated from the hard-wall hQCD model, better fitted to the experimental result, which cannot be achieved by 
other hQCD setups like the Sakai-Sugimoto model~\cite{Harada:2010cn}.

We further integrated out the vector mesons from the HLS Lagrangian to arrive at the ChPT with the coefficients expressed in terms of the hQCD quantities. Through these expressions we found some interesting relations among the coefficients. 
Remarkably, the relation between $L_9$ and $L_{10}$ leads to the vanishing pion axial-vector form factor $F_A=0$.

Finally, we emphasize 
 that our expressions of the coefficients in the chiral effective theories can be quite useful 
for phenomenological applications in estimating quantities of low-energy QCD in terms of the well-established ChPT or HLS-ChPT parameters 
up to ${\cal O}(p^4)$.      
Note that the large amount of coefficients in the chiral effective theory cannot be determined 
phenomenologically at the moment, so 
our approach provides a way to compute the coefficients self-consistently without invoking such a large amount of inputs. 
As one application, 
one might think about the computation of the baryon and baryonic matter properties from mesonic theories, 
in which the $\mathcal{O}(p^4)$ terms are quite essential~\cite{Ma:2012kb,Ma:2012zm,Ma:2013ooa}. 
However, in Refs.~\cite{Ma:2012kb,Ma:2012zm,Ma:2013ooa} the current-quark mass effect is not taken into account 
since the mesonic theory used there is based on the Sakai-Sugimoto model which, at present, cannot go beyond the chiral limit. 
The master formulas presented in this paper make it possible to develop the analysis, as done in other references, by including the explicit breaking effect.

\acknowledgments

We would like to thank H.~Nishihara for carefully reading of the manuscript. The work of Y.-L.M. and M.H. was supported in part by the Grant-in-Aid for Scientific Research
on Innovative Areas (No. 2104) ``Quest on New Hadrons with Variety of Flavors'' from MEXT.
Y.-L.M. was supported in part by the National Science Foundation of China (NSFC) under
Grant No.~10905060.
S.M. was supported in part by the JSPS
Grant-in-Aid for Scientific Research (S) No. 22224003.
The work of M.H. was supported in part by the Grant-in-Aid for Nagoya University Global
COE Program ``Quest for Fundamental Principles in the Universe: From Particles to the Solar
System and the Cosmos'' from MEXT, and the JSPS Grant-in-Aid for Scientific Research
(S) No. 22224003 and (C) No. 24540266.


\begin{thebibliography}{99}


\bibitem{Maldacena:1997re} 
  J.~M.~Maldacena,
  Adv.\ Theor.\ Math.\ Phys.\  {\bf 2}, 231 (1998).


\bibitem{Witten:1998qj}
  E.~Witten,
  Adv.\ Theor.\ Math.\ Phys.\  {\bf 2}, 253 (1998).

\bibitem{Sakai:2004cn} 
  T.~Sakai and S.~Sugimoto,
  Prog.\ Theor.\ Phys.\  {\bf 113}, 843 (2005).

\bibitem{Sakai:2005yt} 
  T.~Sakai and S.~Sugimoto,
  Prog.\ Theor.\ Phys.\  {\bf 114}, 1083 (2005).


\bibitem{Erlich:2005qh} 
  J.~Erlich, E.~Katz, D.~T.~Son, and M.~A.~Stephanov,
  Phys.\ Rev.\ Lett.\  {\bf 95}, 261602 (2005).

\bibitem{Da Rold:2005zs}
  L.~Da Rold and A.~Pomarol,
  Nucl.\ Phys.\  B {\bf 721}, 79 (2005).
  
\bibitem{Kim:2012ey} 
  Y.~Kim, I.~J.~Shin, and T.~Tsukioka,
  Prog.\ Part.\ Nucl.\ Phys.\  {\bf 68}, 55 (2013).

\bibitem{Wei:79}
S.~Weinberg,
Physica A {\bf 96}, 327 (1979).

\bibitem{Bando:1984ej} 
 M.~Bando, T.~Kugo, S.~Uehara, K.~Yamawaki, and T.~Yanagida,
Phys.\ Rev.\ Lett.\  {\bf 54}, 1215 (1985).  

\bibitem{Bando:1987br} 
  M.~Bando, T.~Kugo, and K.~Yamawaki,
  Phys.\ Rep.\  {\bf 164}, 217 (1988).



\bibitem{Harada:2003jx} 
  M.~Harada and K.~Yamawaki,
  Phys.\ Rep.\  {\bf 381}, 1 (2003).



\bibitem{Harada:2006di} 
  M.~Harada, S.~Matsuzaki, and K.~Yamawaki,
Phys.\ Rev.\ D {\bf 74}, 076004 (2006).  


\bibitem{Harada:2010cn}
  M.~Harada, S.~Matsuzaki, and K.~Yamawaki,
  Phys.\ Rev.\  D {\bf 82}, 076010 (2010).


\bibitem{Son:2003et} 
  D.~T.~Son and M.~A.~Stephanov,
  Phys.\ Rev.\ D {\bf 69}, 065020 (2004).


\bibitem{Tanabashi}
M.~Tanabashi,
Phys.\ Lett.\ B {\bf 316}, 534 (1993).


\bibitem{Haba:2010hu} 
  K.~Haba, S.~Matsuzaki, and K.~Yamawaki,
  Phys.\ Rev.\ D {\bf 82}, 055007 (2010);
  M.~Kurachi, S.~Matsuzaki, and K.~Yamawaki,
  Phys.\ Rev.\ D {\bf 88}, 055001 (2013).



\bibitem{Wess:1971yu} 
  J.~Wess and B.~Zumino,
  Phys.\ Lett.\ B {\bf 37}, 95 (1971); 
  E.~Witten,
  Nucl.\ Phys.\ B {\bf 223}, 422 (1983).



\bibitem{Fujiwara:1984mp}
  T.~Fujiwara, T.~Kugo, H.~Terao, S.~Uehara, and K.~Yamawaki,
  Prog.\ Theor.\ Phys.\  {\bf 73}, 926 (1985). 

\bibitem{Kaymakcalan:1984bz} 
  O.~Kaymakcalan and J.~Schechter,
  Phys.\ Rev.\ D {\bf 31}, 1109 (1985).
  


\bibitem{Gasser:1983yg} 
  J.~Gasser and H.~Leutwyler,
  Ann. Phys.\  {\bf 158}, 142 (1984). 
  
\bibitem{Gasser:1984gg} 
  J.~Gasser and H.~Leutwyler,
  Nucl.\ Phys.\ B {\bf 250}, 465 (1985).



\bibitem{Beringer:1900zz} 
  J.~Beringer {\it et al.}  (Particle Data Group Collaboration),
  Phys.\ Rev.\ D {\bf 86}, 010001 (2012).



\bibitem{Hirn:2005nr} 
  J.~Hirn and V.~Sanz,
  J. High Energy Phys. 12 (2005) 030.


\bibitem{Colangelo:2012ipa} 
  P.~Colangelo, J.~J.~Sanz-Cillero, and F.~Zuo,
 J. High Energy Phys. 11 (2012) 012.


\bibitem{Grigoryan:2008up} 
  H.~R.~Grigoryan and A.~V.~Radyushkin,
  Phys.\ Rev.\ D {\bf 77}, 115024 (2008).


\bibitem{Shock:2006qy} 
  J.~P.~Shock and F.~Wu,
  J. High Energy Phys. 08 (2006) 023.


\bibitem{Shock:2006gt} 
  J.~P.~Shock, F.~Wu, Y.~-L.~Wu, and Z.~-F.~Xie,
 J. High Energy Phys. 03 (2007) 064.


\bibitem{Hirn:2005vk} 
  J.~Hirn, N.~Rius, and V.~Sanz,
  Phys.\ Rev.\ D {\bf 73}, 085005 (2006).
  
  

\bibitem{Gubser:1998bc} 
  S.~S.~Gubser, I.~R.~Klebanov, and A.~M.~Polyakov,
  Phys.\ Lett.\ B {\bf 428}, 105 (1998).

\bibitem{DaRold:2005vr} 
  L.~Da Rold and A.~Pomarol,
 J. High Energy Phys. 01 (2006) 157.


\bibitem{Nishihara:2014nva} 
  H.~Nishihara and M.~Harada,
  Phys.\ Rev.\ D {\bf 89}, 076001 (2014).

\bibitem{Ma:2012kb} 
  Y.~-L.~Ma, Y.~Oh, G.~-S.~Yang, M.~Harada, H.~K.~Lee, B.~-Y.~Park, and M.~Rho,
  Phys.\ Rev.\ D {\bf 86}, 074025 (2012).

\bibitem{Ma:2012zm} 
  Y.~-L.~Ma, G.~-S.~Yang, Y.~Oh, and M.~Harada,
  Phys.\ Rev.\ D {\bf 87}, 034023 (2013).

\bibitem{Ma:2013ooa} 
  Y.~-L.~Ma, M.~Harada, H.~K.~Lee, Y.~Oh, B.~-Y.~Park, and M.~Rho,
  Phys.\ Rev.\ D {\bf 88}, 014016 (2013).


  
\end{thebibliography}
\end{document}